\documentclass[12pt]{iopart}

\usepackage{graphicx}   
\usepackage{verbatim}   
\usepackage{color}      
\usepackage{subfigure}  
\usepackage{hyperref}   
\usepackage{multimedia}
\expandafter\let\csname equation*\endcsname\relax
\expandafter\let\csname endequation*\endcsname\relax
\usepackage{amsmath}
\usepackage{mathrsfs}
\usepackage{iopams}
\usepackage{setspace}
\newcommand{\fhn}{FitzHugh-Nagumo system}

\newcommand{\kw}[1]{\frac{1}{#1}}

\newcommand{\arcoth}{\operatorname{arcoth}}
\newcommand{\Det}{\operatorname{Det}}

\begin{document}

\title[Nucleation on curved surfaces]{Nucleation of reaction-diffusion waves on curved surfaces}

\author{Frederike Kneer$^{1,2}$, Eckehard Sch\"oll$^1$, and Markus A. Dahlem$^{3,4}$}, 
\address{$^1$Institut f{\"u}r Theoretische Physik, Technische
 Universit{\"a}t Berlin, Hardenbergstrasse 36, D-10623 Berlin, Germany}
\address{$^2$Institut f{\"u}r Softwaretechnik und Theoretische Informatik Technische
 Universit{\"a}t Berlin, Marchstr. 23, D-10587 Berlin, Germany}
\address{$^3$ Institut f{\"u}r Physik, Humboldt Universit\"at zu Berlin, Robert-Koch-Platz 4,
D-10115 Berlin, Berlin, Germany}
\address{$^4$ Author to whom any correspondence should be addressed: dahlem@physik.hu-berlin.de}

\begin{abstract}
We study reaction-diffusion waves on curved two-dimensional surfaces, and determine the influence of curvature upon the nucleation and propagation of spatially localized waves in an excitable medium modelled by the generic FitzHugh-Nagumo model. We show that the stability of propagating wave segments depends crucially on the curvature of the surface.
As they propagate, they may shrink to the uniform steady state, or expand, depending on
whether they are smaller or larger, respectively, than a critical nucleus. This critical
nucleus for wave propagation is modified by the curvature acting like an effective
space-dependent local spatial coupling, similar to diffuson, thus extending the regime of propagating excitation waves
beyond the excitation threshold of flat surfaces. In particular, a negative gradient
of Gaussian curvature $\Gamma$, as on the outside of a torus surface (positive $\Gamma$), when the
wave segment symmetrically extends into the inside (negative $\Gamma$), allows for stable
propagation of localized wave segments remaining unchanged in size and shape, or
oscillating periodically in size.
\end{abstract}

\maketitle

\section{Introduction}

Wave propagation in excitable extended media described by nonlinear
reaction-diffusion equations has widespread applications in chemistry, biology,
and medicine. A particularly important example are neuronal systems where
spreading depression (SD) waves are associated with a pathological dysfunction
of brain activity that occurs, for instance, during migraine or stroke.
Understanding the effect of internal control mechanisms in neuronal wave
dynamics is of great relevance not only for comprising the functionality of the
human brain \cite{DAH12a}, but also for developing novel future therapies of
pathological states which are connected with cortical spreading depression
waves in migraine aura or stroke \cite{KAR13,DRE11}. There is clinical and
experimental evidence \cite{DAH08d,DAH09a} that spatially localized {\em wave
segments} play a dominant role in these phenomena. In general, spatially
localized wave segments, as they propagate, might shrink, expand, or remain
unchanged in size and shape, in which case they are called {\em particle-like
waves} or {\em dissipative solitons}. Spatially localized wave segments also
represent critical structures which can be stabilized by global feedback
\cite{KRI94,BOD95,SCH97a}. They play an important role for the nucleation of
propagating waves and wave segments in two-dimensional spatial domains.
Generally, waves can be controlled by feedback or {\em closed loop control}.
This is a robust and versatile concept which uses the internal dynamics of the
system to generate a control signal which directs the system towards a desired
dynamics.  A plethora of examples are provided by global or nonlocal and in
some cases time-delayed feedback control of wave propagation in
reaction-diffusion systems \cite{KIM01,ZYK04a,SCH06c,MIK06,DAH08,SCH09c,KYR09}.
On the other hand, the curvature of the medium itself also provides a means of
internal control of the stability, as we will show in this paper.

Most previous studies have focussed on wave propagation in planar spatial
domains, yet there is also a considerable body of work on reaction-diffusion
waves in curved surfaces mostly on spirals and ring waves
\cite{MAS89,ABR90,DAV91a,DAV93,MIK94a,DAV00,DAV00b,DAV02,MAN03b} but not to the
best of our knowlege on nucleation.  The cortex, however, represents a strongly
curved surface. It is the purpose of this paper to study nucleation and
propagation of wave segments on curved two-dimensional surfaces. We will
demonstrate that positive or negative Gaussian curvature of the spatial domain
has a dramatically different effect upon the wave dynamics. 

The paper is organized as follows. In Sect.~\ref{sec:methods} we present the
model. In Sect.~\ref{sec:results} we discuss wave solutions on a torus, which
represents a curved surface on which locally both positive and negative
Gaussian curvature occurs. We consider ring waves, wave segments, and
dissipative solitons (critical nuclei) stabilized by feedback control.
Specifically, we study ring wave break-up, curvature-induced changes of
stability, and curvature-induced stabilisation of wave-segments.  In
Sect.~\ref{sec:conclusion} we draw conclusions.

\begin{figure}
\centering
\includegraphics{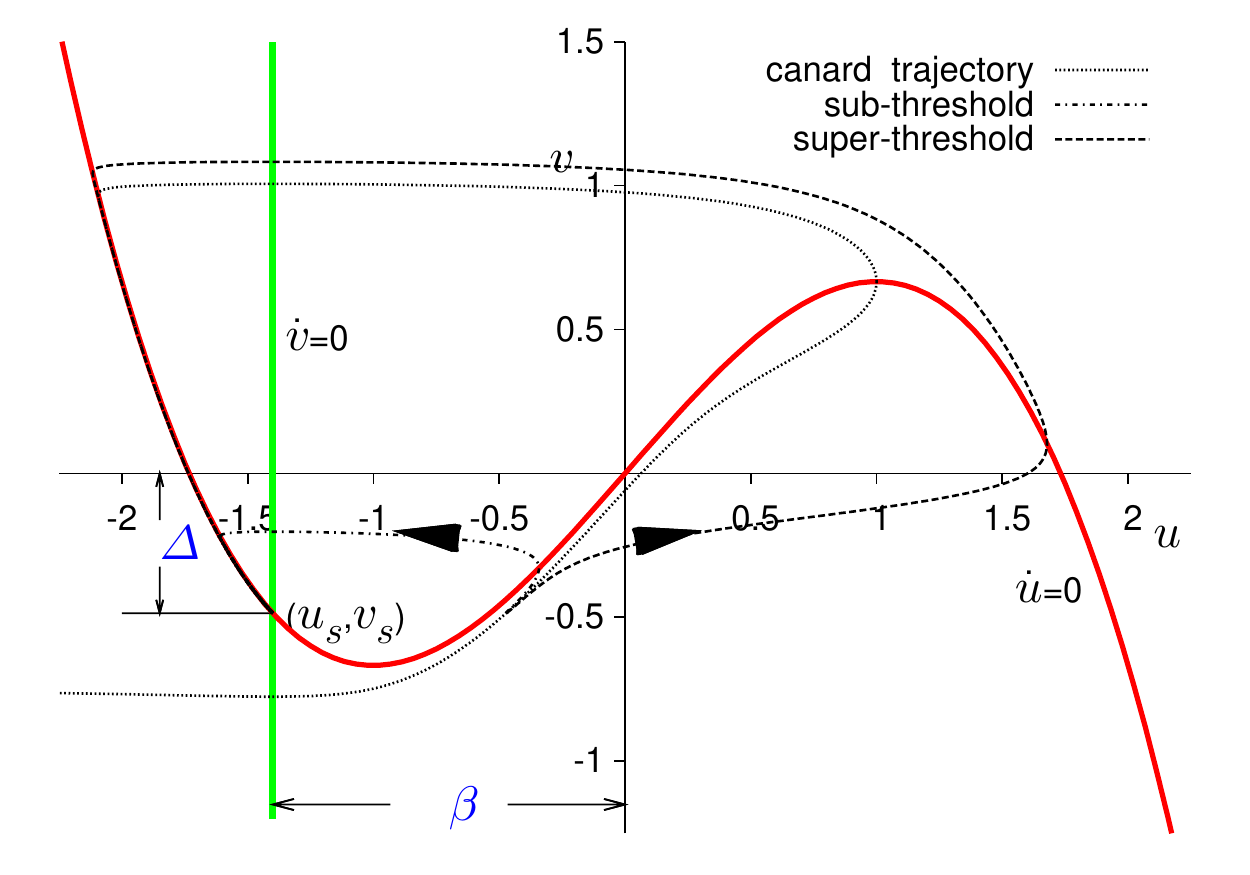}
\caption{The nullclines $\dot u=0$ (solid red) and $\dot v=0$ (solid green) in the phase space of the homogeneous FHN system with $\beta=1.4$. Their intersection at $(u_s,v_s)$ is a stable fixed point. Three trajectories are drawn for  $\varepsilon=0.04$: one canard-like trajectory (dotted), passing  through the maximum of the nullcline $\dot u=0$, and two trajectories starting at $v=v_s$ nearby but on opposite sides of the canard trajectory. They diverge sharply, producing threshold behavior: the dashed and the dash-dotted trajectories represent super-threshold and sub-threshold stimulation, respectively.}
\label{fig:fhn_det_pp_schematic}
\end{figure}

\begin{figure}
 \includegraphics[width=1.0\textwidth]{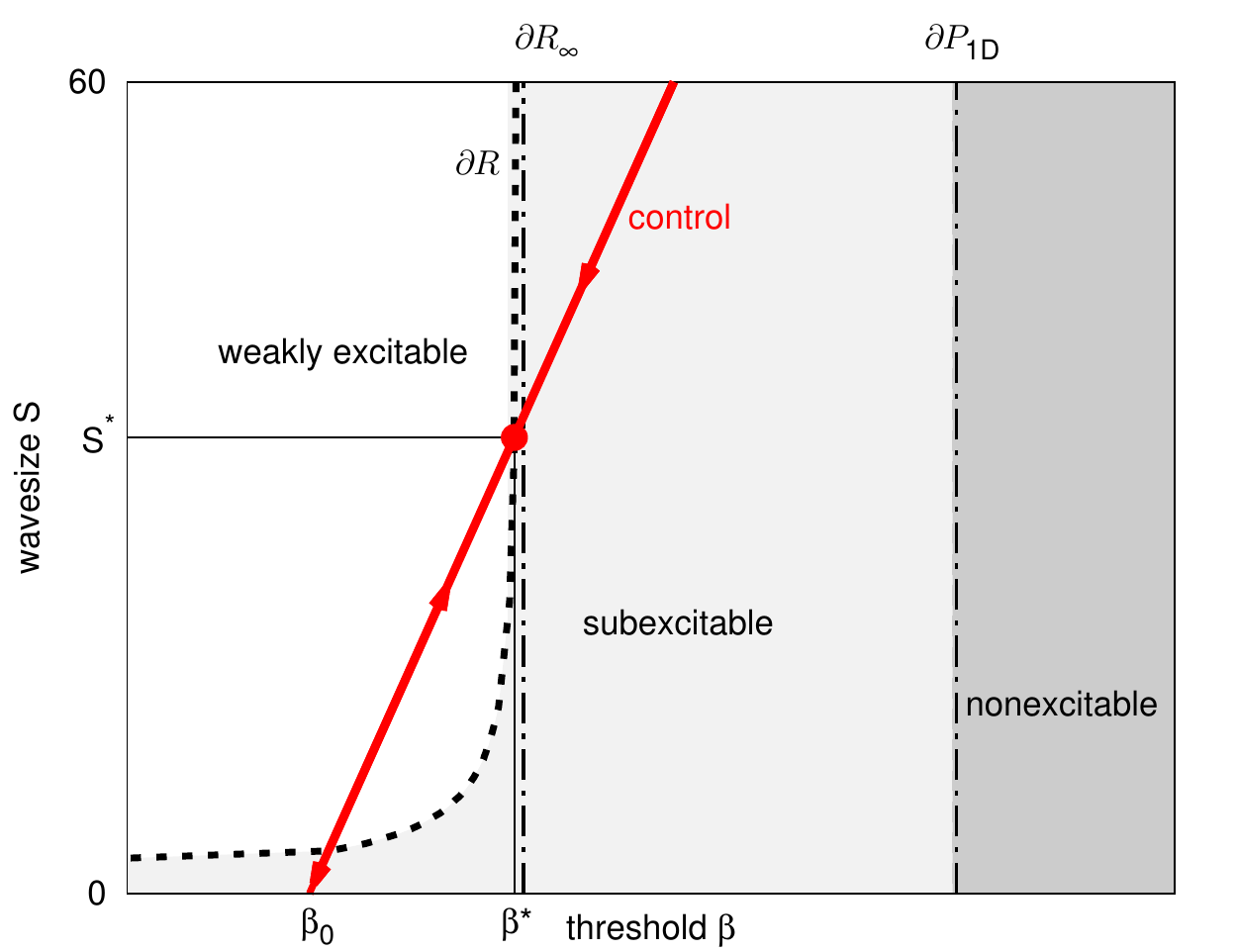}
\caption{Schematic phase diagram of different regimes of the FithHugh-Nagumo model in the plane of wavesize $S$ and threshold $\beta$ for fixed $\varepsilon$: weakly excitable (perturbations grow to spiral waves); subexcitable (perturbations shrink in length); nonexcitable (perturbations shrink in width, no propagation in one dimension). The respective boundaries are marked by $\partial R_{\infty}$ and $\partial P_{1D}$. $\partial R$ denotes the boundary of the critical nucleus of size $S$ below which perturbations shrink. The red solid line marks the control loop, which stabilizes the critical nucleus $(S^*,\beta^*)$ indicated by a red dot.
}\label{fig:bifDiaControl}
\end{figure}

\begin{figure}
\centering
 \includegraphics[width=0.7\textwidth]{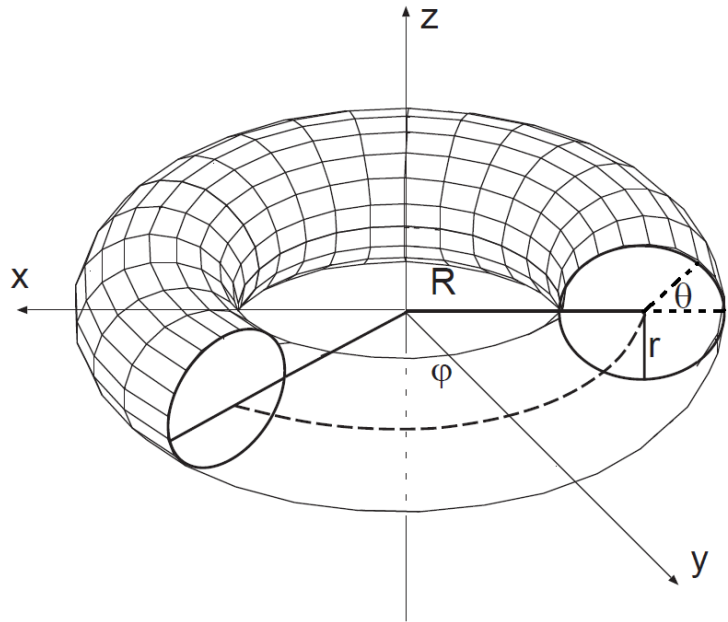}
\caption{Parametrization of a torus by coordinates $(\theta,\varphi)$.}\label{fig:torusCoordiantes}
\end{figure}

\section{Model}
\label{sec:methods}

In this work we use the FitzHugh-Nagumo (FHN) system as a generic model for excitable systems \cite{LIN04}, and study it in two spatial dimensions. It is a 
generalization of the \textit{van der Pol oscillator} \cite{POL26} and also known as 
\textit{Bonhoeffer-van der Pol} oscillator \cite{BON48,BON53}. The model was first suggested by FitzHugh in 1961 \cite{FIT61}, and independently by Nagumo et al. in 1962 \cite{NAG62}. As a 
two-variable simplification of the four-variable \textit{Hodgkin-Huxley} model, which describes the 
propagation of action potentials along the giant squid axon \cite{HOD52}, it describes the response of an excitable nerve membrane to external current stimulation:
\begin{eqnarray} \label{eq:fhn1} 
  \frac{\partial u}{\partial t} &=& 3u - u^3 -v + D \nabla^2u,  \\\label{eq:fhn2}
  \frac{\partial v}{\partial t} &=& \varepsilon( u+ \beta). 
\end{eqnarray} 

The original interpretation of the FHN Eqs.~(\ref{eq:fhn1})-(~\ref{eq:fhn2}) is
based on a single neuron: The variable $u$ models fast changes of the
electrical potential across the membrane of a nerve cell axon (occurring as
spikes in the time series), and $v$ is the recovery variable related to the
gating mechanism of the membrane channels \cite{FIT61}. The small parameter
$\varepsilon \ll 1$ represents the time scale ratio of the two variables.  The
generic dynamical mechanism described by Eqs.~(\ref{eq:fhn1})-(~\ref{eq:fhn2})
was also suggested to describe a fundamentally different ionic excitability of
neuronal tissue that originates from bistable ion homeostasis and is the bais of spreading
depression \cite{DAH04b,HUE14}.

In general, the fast variable $u$ is called the activator variable, whereas the
slow variable $v$ is referred to as inhibitor variable. The diffusion constant
of the activator is $D$, which is chosen as $0.12$ throughout this paper (and
can simply be interpreted as a scaling of space), and inhibitor diffusion is
assumed to be slow, and hence negligible.  The threshold parameter $\beta$
determines whether the systems is excitable ($\beta>1$) or exhibits
self-sustained perodic oscillations $(\beta<1)$. In the following we consider
$\beta$ in the excitable regime.\\

Fig.~\ref{fig:fhn_det_pp_schematic} shows a schematic phase portrait of a spatially homogeneous system in the excitable regime ($\beta>1$) with the cubic activator nullcline and the vertical inhibitor nullcline (solid red and green lines). The system has a single fixed point $(u_s, v_s)$, which is stable for $\beta>1$ and located on the left branch of the cubic nullcline. At $\beta=1$ a supercritical Hopf bifurcation of a limit cycle occurs, and the fixed point becomes unstable and is shifted to the middle branch of the nullcline for $\beta<1$. 

The excitable behavior of the system is crucially determined by the cubic non-linearity of the activator equation and the separation of time-scales between the two variables: When the system is perturbed by a sufficiently large (super-threshold) external stimulus, which can be regarded as setting the initial condition, the system undergoes a large excursion in phase space (spiking). Starting from its initial condition, the system performs, due to the strong timescale separation $\varepsilon\ll1$, a fast transition to the stable right branch of the activator nullcline. After that, it travels slowly upwards approximately along this nullcline, until the phase points jumps back to the left branch, and returns, along the left branch of the nullcline, slowly downwards to the fixed point (recovery phase). Without further external stimulation the system remains in the stable fixed point (rest state). 

The threshold-like behavior of the \fhn{} is associated with the \textit{canard-like trajectory} 
(dotted black line in Fig.(~\ref{fig:fhn_det_pp_schematic})), which is the trajectory passing through 
the local maximum of the cubic nullcline, and which is often referred to as the threshold of the 
\fhn{}. The region around the canard-like trajectory is extremely sensitive to initial conditions: For 
initial conditions only slightly below the canard-like trajectory the systems will perform a large 
excursion in phase space, whereas for initial conditions only slightly above the canard-like trajectory the excursion will be small (sub-threshold excitation). In principle, the transition from small to large amplitude excitation is continuous; in fact, however, phase space excursions of intermediate amplitude are very rare. Correspondingly, small sub-threshold stimulations (dash-dotted black line in Fig.(~\ref{fig:fhn_det_pp_schematic})) will result in fast relaxation, while super-threshold stimulations (dashed black line in Fig.(~\ref{fig:fhn_det_pp_schematic})) induce a full excursion in phase space, corresponding to a characteristic spike in the time evolution of the $u$-variable. 

In the following we consider spatially inhomogeneous solutions, i.e., waves or wave segments in two spatial dimensions (2D). Depending upon the set of parameters $(\varepsilon, \beta)$ there exist different wave solutions \cite{WIN91,DAH09a}. Here we focus on localized wave segments which may either shrink
or expand, as they propagate, or in the limit case, remain unchanged in size and shape, in which case they are called {\em particle-like waves} or {\em dissipative solitons}.

For each set of parameters $(\varepsilon, \beta)$ with $\beta<\beta_{\partial R_{\infty}}$ (or $\varepsilon<\varepsilon_{\partial R_{\infty}}$, respectively) there exists a
localized wave solution (wave segment), 
which represents a critical spatio-temporal structure, i.e., a {\em particle-like wave}, or {\em dissipative soliton}.

As we are interested in stationary propagating wave segments, Eqs.~(~\ref{eq:fhn1})-(~\ref{eq:fhn2}) can be written as
\begin{eqnarray} \label{eq:fhn1comoving} c\frac{\partial u}{\partial
  \xi} &=& 3u - u^3 -v + D \nabla_{\xi}^2u,  \\\label{eq:fhn2comoving}
  c\frac{\partial v}{\partial \xi} &=& \varepsilon( u+ \beta), \end{eqnarray} 
where $\xi=x+ct$ is the co-moving coordinate with propagation velocity $c$.\\
In a co-moving frame, the localized critical structure is related to a saddle-point with a single 
unstable eigenvector (one-dimensional unstable manifold) in phase space. The curve representing this solution in a parameter plane of the bifurcation diagram is called the rotor boundary $\partial R$. In phase space, the stable manifold of states on $\partial R$ separates the attractor of a spiral wave (spatially
non-confined) and the stable, spatially uniform steady state. Thus, when the
form of this dissipative soliton is perturbed, it either grows to a spiral
wave, or shrinks and disappears. Perturbations above the critical size of the dissipative soliton grow, while perturbations below that critical nucleus shrink to the stable uniform state, i.e., the wave segments retracts.   The internal cortical control of such dissipative solitons may be viewed as a strategy of the cortex to avoid re-entrant spiral waves, e.g., in migraine. 
Changing  the nucleation size of this critical structure changes the
susceptibility to pathological conditions such as spreading depression.\\

In Fig.~\ref{fig:bifDiaControl}, we show the rotor boundary $\partial R$ (black dashed) in a schematic bifurcation diagram of wavesize $S$ as a function of the threshold parameter $\beta$. It separates the weakly excitable parameter regime (perturbations grow to a spiral wave) from the subexcitable parameter regime (wave segments in 2D below the critical size shrink in length, while in spatially one-dimensional (1D) systems wave propagation is stable). There exists another boundary $\partial R_{\infty}$ (dash-dotted vertical line), independent of size $S$, to the right of which all wave segments retract (corresponding to infinitely large critical size).
Furthermore, the propagation boundary $\partial P_{1D}$ is shown, which separates the subexcitable parameter regime from the nonexcitable regime. In the nonexcitable parameter regime, perturbations shrink also in width and wave segments collapse, i.e., even in spatially one-dimensional systems no wave propagation is possible.\\

At this point we would like to remark that the critical nucleus (dissipative soliton), which has the dynamic signature of a saddle-point, can be stabilized by an internal feedback control loop which controls the excitation threshold $\beta$ in
Eq.~(~\ref{eq:fhn2}) that is,
\begin{eqnarray}
\label{tp-b7-eq:gconstraint}
\beta(t) &=&  \beta_0 + K S(t) 
\end{eqnarray} 
where $K$ is the control strength, and the size of the wave segment $S$ represents a measure of the active area occupied by the wave segment. \\

Eq.~(\ref{tp-b7-eq:gconstraint}) defines a control line (red solid line with arrows) in the $(\beta, S)$ phase diagram in Fig.~\ref{fig:bifDiaControl}. As the temporal evolution of a perturbation in a controlled system is confined to the control line, it asymptotically approaches a stable wave segment $(\beta^*,S^*)$ (if perturbed with convenient initial conditions). This follows from Fig.~\ref{fig:bifDiaControl} since wave segments above $\partial R$, i.e., $\beta < \beta^*$, grow in size, while wave segments below $\partial R$, i.e., $\beta > \beta^*$, shrink.\\

The size of the wave segment $S$ can be calculated via an integral over the excited area, e.g.,
\begin{eqnarray}
 S(t)&=& \int_{\mathbb{R}^2}u({\bf r},t)\,d^2{\bf r}.
\end{eqnarray}
If the stabilized solution $(\beta^*, S^*)$ is reached, $S(t)$ becomes stationary. Thus, although the control is invasive, it does
not produce new solution branches, and the stabilized critical nuclei of the controlled system are unstable solutions of the system without control at $\beta=\beta(t\longrightarrow\infty)$. Hence, it is adequate to define the size of the wave segment as the area where the activator concentration $u({\bf r},t)$ is larger than zero
\begin{eqnarray}
 S(t)&=&\int_{\mathbb{R}^2}\Theta(u({\bf r},t))\,d^2{\bf r}
 \label{eq:S}
\end{eqnarray}
with the Heaviside function $\Theta$.\\

In the following simulations we will apply an internal feedback mechanism as in Eq.~(\ref{tp-b7-eq:gconstraint}) with the wavesize $S(t)$ from Eq.~(\ref{eq:S}) to stabilize the critical nucleus.
Furthermore, we will study the influence of the curvature of the excitable medium on
the stability of localized waves. Hence the Laplace operator $\nabla^2$ must be
replaced with the Laplace-Beltrami operator $\Delta_{LB}$ \cite{KUE05} for surfaces given in curvilinear coordinates $\alpha^i$, $i=1,2$,:
\begin{eqnarray}
\label{eq:LB}
\Delta_{LB} &=& \sum^{2}_{i,j=1} g^{-\frac{1}{2}}\frac{\partial}{\partial\alpha^i}\left(g^{\frac{1}{2}}g^{ij}\frac{\partial }{\partial\alpha^j}\right)
\end{eqnarray}
where $g=\Det\textbf{G}$ and $\textbf{G}$ with the matrix elements $g^{ij}$ is the metric tensor of the parametrization, see Appendix.\\
The surface of a torus in the Euclidian space $\mathbb{R}^3$ can be described by the parametrization $(\theta,\varphi)$ of the position vectors

\begin{eqnarray}
(\theta,\varphi)
&\mapsto&
\left(
\begin{array}{c}
 (R+r\cos\theta)\cos\varphi\\(R+r\cos\theta)\sin\varphi\\r\sin\theta
\end{array}
\right)
=
\left(
\begin{array}{c}
 x\\y\\z
\end{array}
\right).
\end{eqnarray}
The geometrical meaning of the major curvature radius $R$ and the minor curvature radius $r$ as well as the angles $\theta$ and $\varphi$ is visualized in Fig.~\ref{fig:torusCoordiantes}.
The Laplace-Beltrami operator in torus coordinates reads
\begin{eqnarray}
\boldsymbol{\Delta}_{LB}&=&-\frac{\sin\theta}{r(R+r\cos\theta)}\frac{\partial}{\partial\theta}+\frac{1}{r^2}\frac{\partial^2}{\partial\theta^2}+\frac{1}{(R+r\cos\theta)^2}\frac{\partial^2}{\partial\varphi^2}.
\label{eq:Laplacebeltramig}
\end{eqnarray}
We investigate sections of a torus with von Neumann boundary conditions
(no flux boundary) on the equatorial section (at $\theta=0$ and $\theta=\pi$) and periodic boundary conditions in the direction of the azimuthal angle $\varphi$. This restricts all traveling wave solutions as they have
to obey the symmetries defined by these boundary conditions, i.e., the center of mass of the critical nucleus is pinned either on the outside or inside of the torus.\\

The FHN model with diffusion and internal feedback control Eq~(\ref{tp-b7-eq:gconstraint}) including the wavesize $S(t)$ calculated with Eq~(\ref{eq:S})
\begin{eqnarray}
 \frac{\partial u}{\partial t}&=& 3 u - u^3 -v + D\Delta_{LB}u,\\
 \frac{\partial v}{\partial t}&=&\varepsilon(u + \beta_0 + KS(t))
\end{eqnarray}
is numerically solved using the explicit Euler method. With the discretization of space and time 
\begin{eqnarray}
 x_j:=x_0+j\delta x,   j=0,1,....,J\nonumber\\
 t_n:=t_0+n\delta t,   n=0,1,....,N,\nonumber
\end{eqnarray}
where $x$ stands for $\theta$ or $\varphi$, respectively, the time derivative is calculated as
\begin{eqnarray}
 u_{j}^{n+1}&=&u_{j}^{n}+(3u_{j}^{n}-(u_{j}^{n})^3-v_{j}^{n}+D \Delta_{LB}u_{j}^{n})\delta t,\\
 v_{j}^{n+1}&=&v_j^n+\varepsilon (u_{j}^{n}+ \beta_0 +KS(t))\delta t.
\end{eqnarray}
with $u^n_j:=u(t_n,x_j)$. The Laplace-Beltrami operator Eq.~(\ref{eq:LB}) contains derivatives of first and second order with respect to $\theta$ and $\varphi$. The derivative of first order is solved using a forward-backward-Euler-algorithm
\begin{eqnarray}
 \frac{\partial u}{\partial\theta}\approx\frac{u_{j+1}^n-u_{j-1}^n}{2\delta\theta}.
\end{eqnarray}
The derivatives of second order are solved using the Euler method, first backward followed by forward
\begin{eqnarray}
 \frac{\partial^2}{\partial\theta^2}\approx\frac{u_{j+1}^n-2u_j^n+u_{j-1}}{(\delta\theta)^2}.
\end{eqnarray}
As initial condition, the activator concentration $u$ is set equal to $u_s+2$ (which corresponds to a supra-threshold excitation in
Fig.~\ref{fig:fhn_det_pp_schematic}) in a rectangular area, and the inhibitor concentration $v$ is set equal to $v_s+1.5$ in a rectangular area shifted relative to the activator stimulus, in order to determine the propagation direction of the wave segment. Outside the rectangle, the initial condition is $u=u_s$, $v=v_s$, with the activator and inhibitor fixed point values $u_s, v_s$. In order to numerically obtain critical nuclei of smaller or larger size, the size of the rectangular area is varied.

\section{Results}
\label{sec:results}
\subsection{Overview of wave solutions on a torus}

\begin{figure}[b!]
\centering
\includegraphics[width=0.8\textwidth]{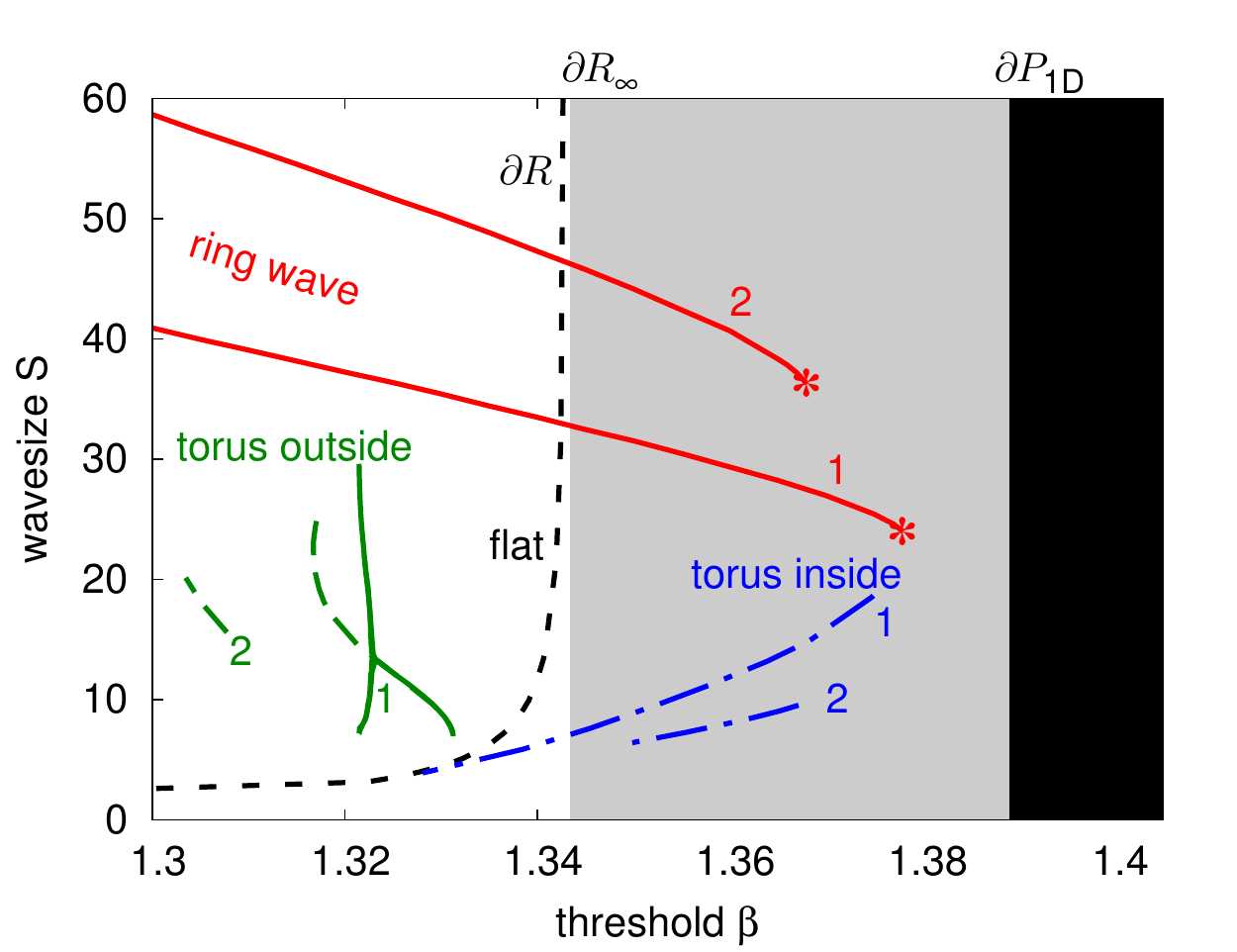}
\caption{Bifurcation diagram of wavesize $S$ as a function of threshold parameter $\beta$ computed from Eqs.~(\ref{eq:fhn1}),(\ref{eq:fhn2}) with $D=0.12$ and $\varepsilon=0.36$; critical nucleus on a flat surface (black dashed line); 1: solutions on a torus with minor curvature radius $r=\frac{20}{2\pi}$ and major curvature radius $R=\frac{80}{2\pi}$; 2: solutions on a torus with minor curvature radius $r=\frac{20}{2\pi}$ and major curvature radius $R=\frac{40}{2\pi}$; stable ring wave solutions (red solid lines) with points of excitation block, i.e., propagation suppression (red asterisks); unstable inside critical nucleus (blue dash-dotted lines); unstable outside critical nucleus (green dashed lines); stable stationary and stable oscillating localized wave segment on the torus outside (green solid lines). Feedback Eq.~(\ref{tp-b7-eq:gconstraint}) is applied to stabilize the states on the dashed and dash-dotted curves.}
\label{fig:bifDiaResults}
\end{figure}

\begin{figure}[b!]
\centering
\includegraphics[width=0.8\textwidth]{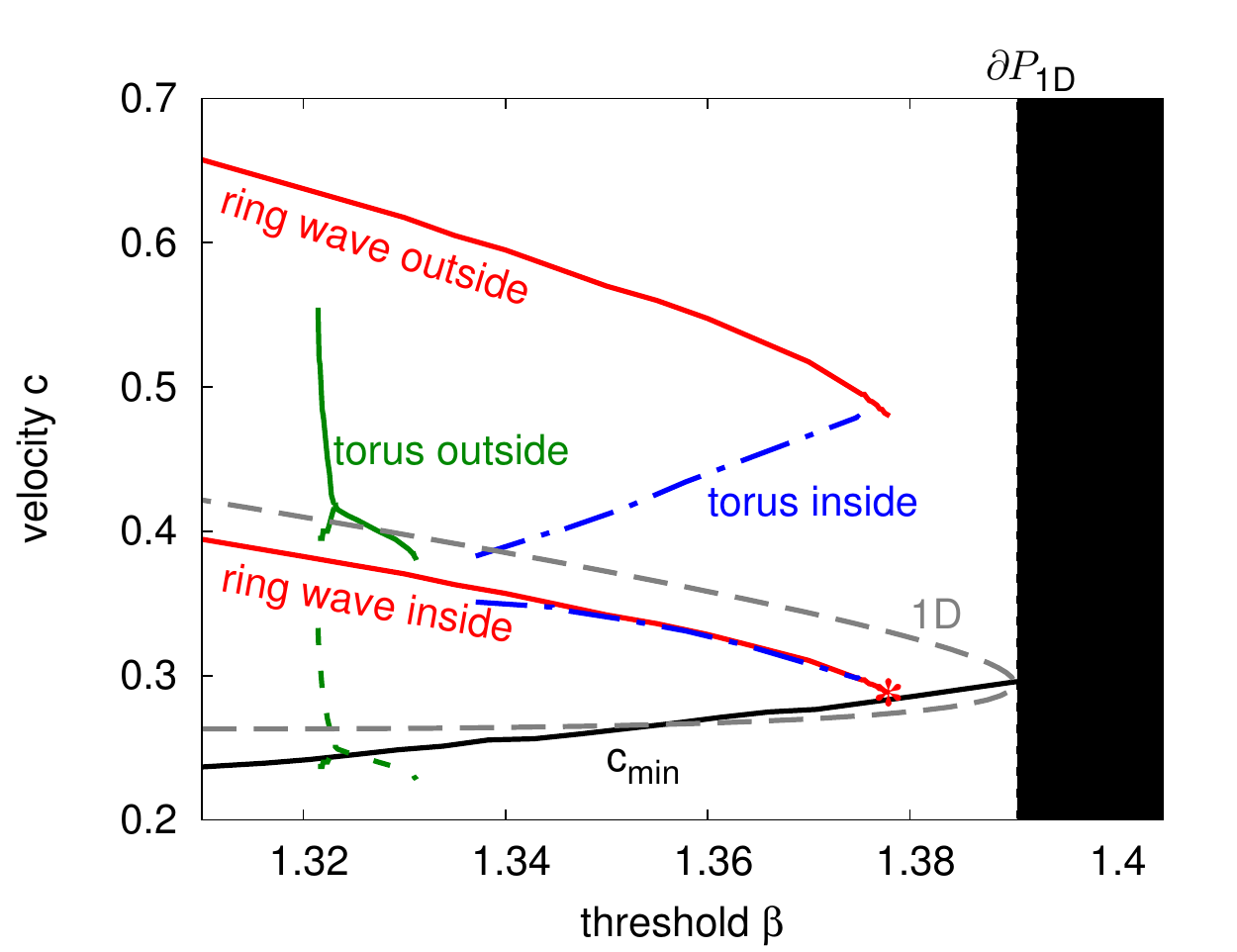}
\caption{Bifurcation diagram of propagation velocity $c$ as a function of threshold $\beta$ computed from Eqs.~(\ref{eq:fhn1}),(\ref{eq:fhn2}) with $D=0.12$ and $\varepsilon=0.36$ on a torus with minor curvature radius $r=\frac{20}{2\pi}$ and major curvature radius $R=\frac{80}{2\pi}$; propagation velocity of stable ring wave solution (red solid lines) on the torus inside (lower line) and torus outside (upper line), point of excitation block (red asterisk); propagation velocity of inside critical nucleus (blue dash-dotted lines) at center of mass (lower line) and open ends (upper line); propagation velocity of stable stationary and stable oscillating localized outside wave segment (green solid lines) at center of mass (upper line) and at open ends (lower line); hypothetical propagation velocity of outside stable wave segments (green dashed line) on torus inside; minimum propagation velocity (black solid line) calculated from Eq.~(\ref{eq:cmin}) with $c_{cr}$ and $\varepsilon_{cr}$ computed from Eqs.(~\ref{eq:
fhn1}),(~\ref{eq:fhn2}) in one spatial dimension; propagation velocity of the stable and unstable wave solution in 1D (grey dashed). Feedback Eq.~(\ref{tp-b7-eq:gconstraint}) is applied to stabilize the states on the blue dash-dotted curves.}
\label{fig:velocity}
\end{figure}

The main results are twofold. First, by investigating excitation waves on a torus,
we show that the Gaussian curvature of the excitable medium changes the
nucleation threshold in a systematic way. Second, and more surprisingly, we
observe that a curved medium can even induce a change of stability.  Unstable
critical nuclei are transformed into stable propagating localized wave segments.

We analyze the nucleation of excitation waves in reaction-diffusion media
on curved 2D surfaces, specifically on tori. A torus has positive and negative Gaussian curvature on the outside ($\theta\!=\!0$) and the inside ($\theta\!=\!\pi$), respectively, and a continuous transition in between
with vanishing Gaussian curvature on the top ($\theta\!=\!\frac{\pi}{2}$) and bottom ($\theta\!=\!\frac{3\pi}{2}$), see Fig.~\ref{fig:torusCoordiantes}. In general, a torus has, in contrast to a
sphere, not only locally varying and even negative Gaussian curvature, but a torus also admits a global isothermal coordinate system, called toroidal coordinates
($(\theta_i, \tilde{\varphi})$, see Appendix), that is, coordinates  where
the metric is locally conformal to the Euclidean metric. Therefore an intuitive
understanding of some of our results can be based on the particularly simple form of
the Laplace-Beltrami operator, the ``diffusion operator'', in these
coordinates.

On tori, a stable solution besides the spatially homogeneous steady state are
ring-shaped localized traveling wave solutions.  Ring-shaped traveling waves have
been analyzed, and in particular their critical properties have been discussed,
namely, wave fronts with sufficiently large geodesic curvature break up on
the torus inside \cite{DAV03}. We reconsider these travelling waves in order to
compare them with the dynamics of critical nuclei. The stable manifold
of a critical nucleus separates the attractor of a ring-shaped
traveling wave and the spatially uniform steady state.

We restrict our study to nucleation of waves propagating strictly in the
direction of the azimuthal angle $\varphi$ (see Sect.(~\ref{sec:methods})) and,
furthermore, the center of mass of the nucleation is pinned either on the
outside or inside of the torus, i.e., the locations where the extreme values of
the Gaussian curvature occur. In the following, we will simply refer to these
solutions as inside or outside critical nuclei or, if stabilized, inside or
outside traveling wave segments. These solutions are the symmetric solutions
with respect to the equatorial section of the torus. Note that there exist also
asymmetric solutions on tori that we have not studied.  Furthermore, note that
these two symmetric critical nuclei on the outside and inside assume the shape
of a localized wave segment where the open ends extend in the direction of
$\theta$ (perpendicular to the propagation direction), that is, into regions of
decreasing and increasing Gaussian curvature, respectively. Our results are
mainly explained by this gradient in the Gaussian curvature and not by the
absolute value of the Gaussian curvature.

The results are displayed in two bifurcation diagrams.
First, the same bifurcation diagram as already introduced in Sect.~\ref{sec:methods} to define
the regimes of excitability (weakly, sub-, and nonexcitable, see
Fig.~\ref{fig:bifDiaControl} in Sect.~\ref{sec:methods}) is shown in
Fig~\ref{fig:bifDiaResults}. The size $S$ of the critical nucleus, see
Eq.~(\ref{eq:S}), is plotted versus the threshold parameter $\beta$ of the local
dynamics of the FHN system Eq.~(\ref{eq:fhn1})-(\ref{eq:fhn2}).  The reference branch of the critical nucleus from simulation on a
flat medium is now labeled ``flat'' in
Fig~\ref{fig:bifDiaResults} (black dashed). It  separates the weakly excitable regime (to the left, decreasing
$\beta$) from the subexcitable regime (to the right, increasing $\beta$), which
ends at $\beta=\partial P_{1D}$, where the nonexcitable regime
is reached. 

In Fig~\ref{fig:bifDiaResults}, we show further solution branches simulated on two different tori. The torus labeled 1 has lower absolute values of Gaussian curvature than the torus labeled 2, since the latter torus has a
two times smaller value of its major curvature radius $R$.
For each torus, we have a branch of the ring-shaped traveling wave solution (red solid). Furthermore, for each torus, we have one branch of the inside (blue dash-dotted) and outside (green dashed) critical nucleus. The states on the curves of the critical nucleus (dashed or dash-dotted) are stabilized by applying an appropriate global feedback Eq.~(\ref{tp-b7-eq:gconstraint}) with suitably chosen $\beta_0$ and $K$ such that the respective state $(\beta^*, S^*)$ is at the intersection with the line given by Eq.~(\ref{tp-b7-eq:gconstraint}).
In addition, on the torus outside, we find stable wave segments and stable oscillating waves (green solid), see Sect.~\ref{sec:resultsStabilisation} below.

Second, Fig.~\ref{fig:velocity} is a bifurcation diagram, where the propagation velocity $c$, see Eq.~(\ref{eq:fhn1comoving})-(\ref{eq:fhn2comoving}), is plotted versus the threshold parameter $\beta$. The reference branches are on the one hand the propagation velocities of the stable fast and the unstable slow wave solution in spatially one-dimensional systems with diffusion in one spatial direction only, \cite{KRU97}), (grey dashed), and,  on the other hand, a critical velocity $c_{min}$ (black solid), below which stable wave propagation cannot be obtained \cite{DAV03}.

In Fig.~\ref{fig:velocity}, we show further solution branches simulated on the less curved torus (torus 1). Two branches show the propagation velocity $c$ in azimuthal ($\varphi$) direction of the ring-shaped traveling wave solution (red solid), the lower one is the propagation velocity on the torus inside, the upper one the propagation velocity on the torus outside. Furthermore, for the inside critical nucleus (blue dash-dotted), we display the propagation velocity in azimuthal ($\varphi$) direction at the center of mass (lower line) and at the open ends (upper line), where the open ends are defined as the most distant lateral location where the activator concentration $u$ equals zero. 

For the outside stable wave segments, we show the propagation velocity $c$ at the center of mass (green solid) and, after the bifurcation into two branches with decreasing threshold parameter $\beta$, the maximum and minimum propagation velocity of the stable oscillating wave segment.
In addition, for the outside critical nucleus (with propagation velocity $c_o$), we plot a ``hypothetical'' branch (green dashed) that shows the 
propagation velocity $c_i=c_o \frac{R-r}{R+r}$ which a point of this wave segment would have on the torus inside 
if it existed there.

\subsection{Ring wave break-up at saddle-node bifurcation}
\label{sec:breakup}

\begin{figure}
\centering
\subfigure{\includegraphics[width=0.8\textwidth]{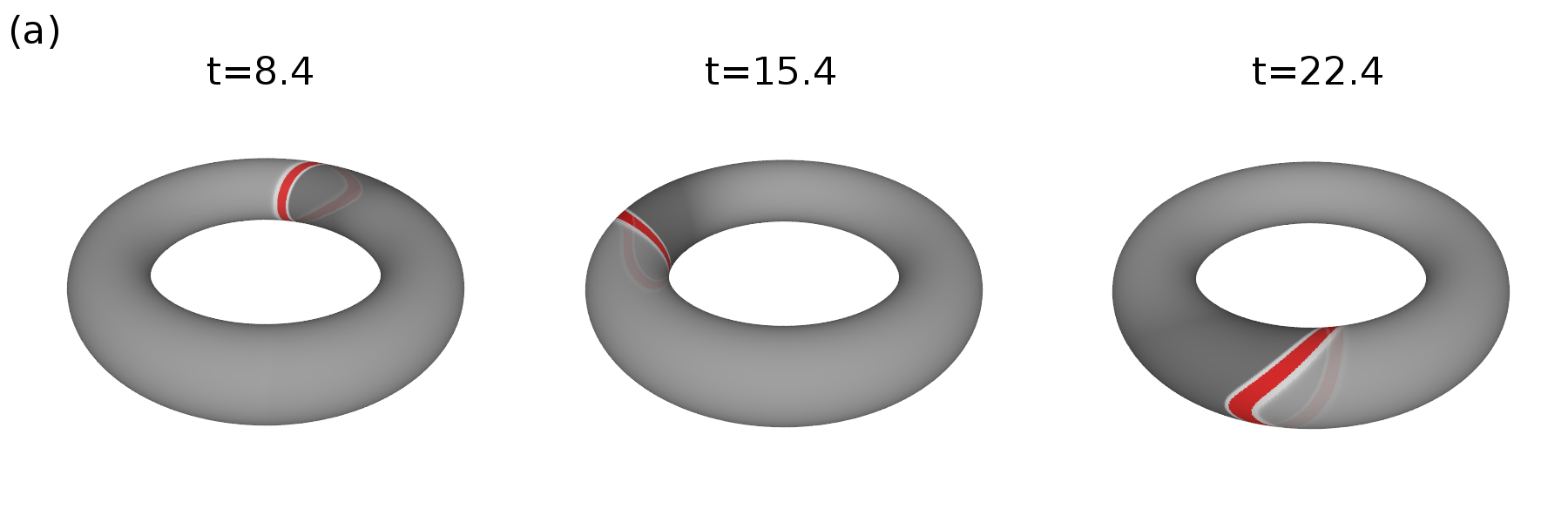}}
\subfigure{\includegraphics[width=0.8\textwidth]{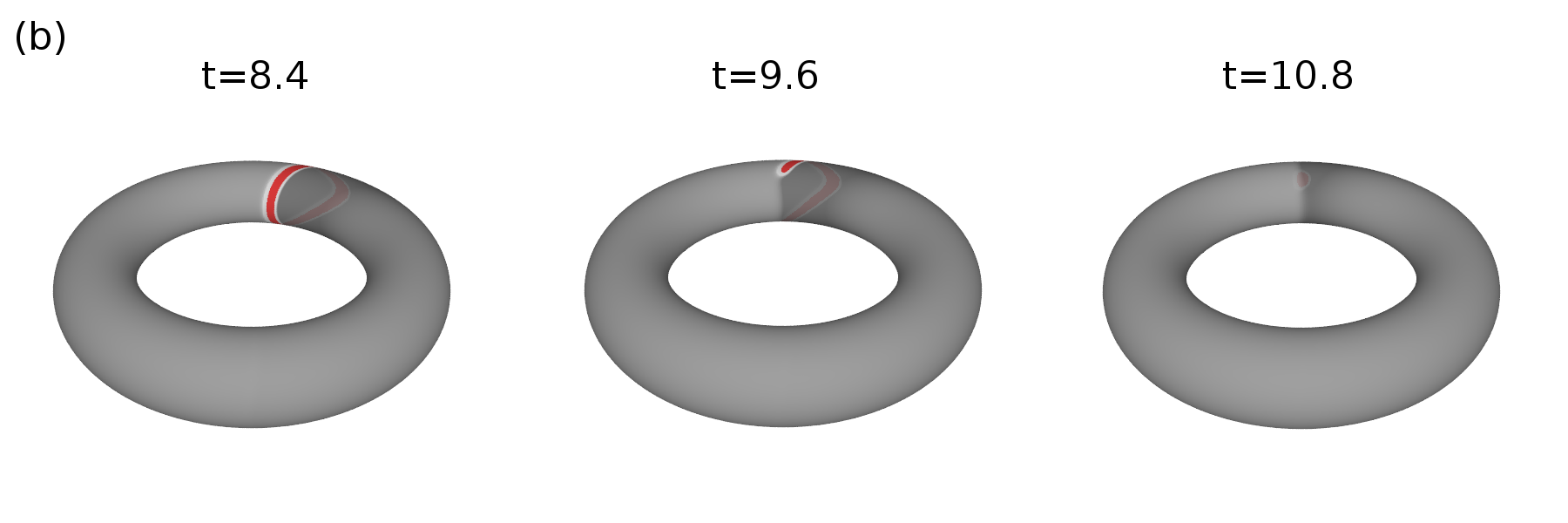}}
\caption{Snapshots of ring waves propagating counter-clockwise on a torus with minor curvature radius $r=\frac{20}{2\pi}$ and major curvature radius $R=\frac{80}{2\pi}$  computed from Eqs.\,(\ref{eq:fhn1})-(\ref{eq:fhn2}) with $D=0.12$ and $\varepsilon=0.36$. (a) Stable ring wave, $\beta=1.378$. (b) Ring-wave break-up, $\beta=1.379.$. 
Movies  are available from  (a) http://www.youtube.com/watch?v=mljU23mrCUg  (b) https://www.youtube.com/watch?v=VaCZQzgOTZ4}
\label{fig:4.6ring}
\end{figure}

\begin{figure}
\centering
 \includegraphics[width=.5\textwidth]{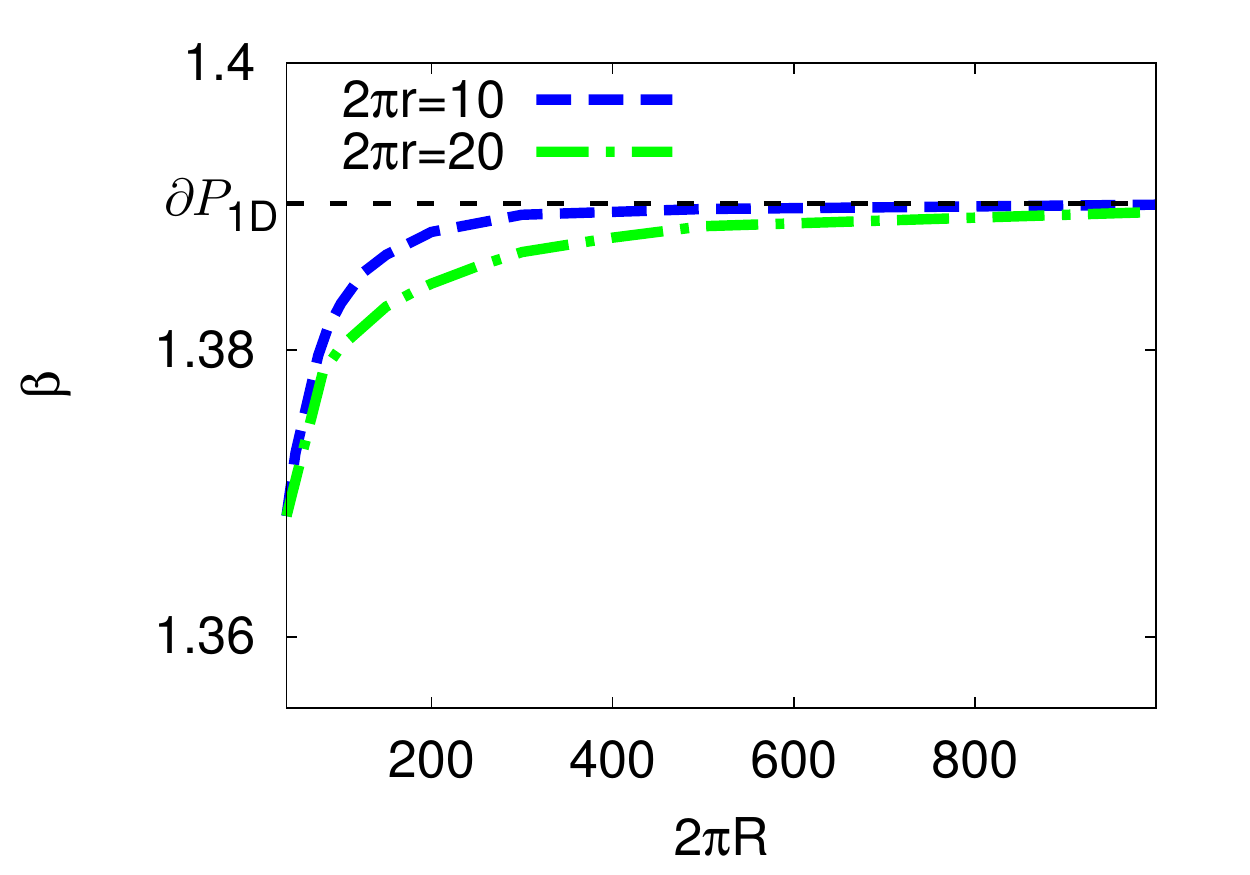}
 \caption{Excitation block (propagation failure) of ring-shaped traveling waves on tori in the ($R,\beta$) parameter space. No propagation is possible above the critical curves for different $r$; $\partial P_{1D}$ denotes the 1D propagation boundary.}
 \label{fig:betaR}
\end{figure}

First we focus on the break-up of ring-shaped traveling waves on tori \cite{DAV03}. The ring-shaped traveling wave solution is a stable 
solution of Eqs.~(\ref{eq:fhn1})-(\ref{eq:fhn2}) shown in Fig.~\ref{fig:4.6ring}. Thus the ring waves can be 
concieved as homoclinic solutions of the related ordinary differential 
equations~(\ref{eq:fhn1comoving}),(\ref{eq:fhn2comoving}) in the co-moving frame.\\
Ring waves have negative geodesic curvature on the torus inside and positive geodesic curvature on the torus outside, see Fig.~\ref{fig:4.6ring}(a).
Thus, compared to 1D pulses (or infinitely extended wavefronts on a flat surface, respectively), ring waves propagate more slowly on the torus inside and faster on the torus outside, see Fig.~\ref{fig:velocity} (red solid).\\
If the propagation velocity falls below a critical value $c_{min}$, the ring wave breaks up on the torus inside \cite{DAV03}, see Fig.~\ref{fig:4.6ring}(b). This excitation block is marked by an asterisk in Figs.~\ref{fig:bifDiaResults} and ~\ref{fig:velocity}. Below the minimal velocity $c_{min}$, stable wave propagation cannot be obtained.\\
For 1D waves, it is known that the propagation failure is due to the coalescence of the homoclinic orbits of the fast and the slow wave (pulse) solution of the ODE problem Eq.~(\ref{eq:fhn1comoving})-(\ref{eq:fhn2comoving}) \cite{KRU97}. In the related PDE problem Eqs.~(\ref{eq:fhn1})-(\ref{eq:fhn2}), the propagation boundary is a saddle-node bifurcation point of the stable fast wave branch and the unstable slow wave branch.\\
In Fig.~\ref{fig:velocity}, the fast wave branch and the slow wave branch of 1D traveling wave solutions are shown (upper and lower dashed grey line). At the propagation boundary $\partial P_{1D}$, they meet in a saddle-node bifurcation.\\
Also in curved 2D media, the propagation failure is due to a saddle-node bifurcation, where the fast wave branch collides with the slow wave branch in a saddle-node bifurcation. In the 1D limit $R\longrightarrow\infty$, the threshold $\beta$, at which the ring waves break up, converges to the propagation boundary $\partial P_{1D}$. This is shown in Fig.~\ref{fig:betaR}, where the lines show the propagation failure in the ($R,\beta$) parameter space on two different tori;
the upper line (dashed blue) is computed on a less curved torus with lower absolute values of Gaussian curvature compared to the lower line (dash-dotted green).\\
The minimum velocity $c_{min}$, below which stable wave propagation is not possible, can be calculated as (see Appendix)
\begin{eqnarray}
 c_{min}&=&\frac{\varepsilon}{\varepsilon_{cr}}c_{cr},
 \label{eq:cmin}
\end{eqnarray}
where $\varepsilon_{cr}$ is the critical time separation parameter, where the homoclinic orbits of the 1D fast and slow wave (pulse) solution of the ODE Eq.~(\ref{eq:fhn1comoving})-(\ref{eq:fhn2comoving}) coincide, and $c_{cr}$ is the corresponding critical velocity. For the line $c_{min}$ shown in Fig.~\ref{fig:velocity}, $\varepsilon_{cr}$ and $c_{cr}$ are computed with AUTO from Eqs.~(\ref{eq:fhn1})-(\ref{eq:fhn2}) in one spatial dimension. AUTO is a software tool for continuation and bifurcation problems in ordinary differential equations.
Here, homoclinic solutions of the FHN model in a co-moving frame Eqs.~(\ref{eq:fhn1comoving})-~(\ref{eq:fhn2comoving}) in one spatial dimension are continued in the threshold parameter $\beta$.\\
The propagation velocity $c$ of ring waves is affected by both the parameters of the local dynamics ($\varepsilon$ and $\beta$) and the Gaussian curvature $\Gamma$ of the torus. An increase of $\beta$ or $\varepsilon$ causes a deceleration of the ring wave on the torus inside. Also an increase of the Gaussian curvature $\Gamma$ of a torus causes an increase of the absolute values of the geodesic curvature of the ring wave, which results in a deceleration of the ring wave on the torus inside.\\
Thus, on more strongly curved tori (smaller $R$), the ring wave breaks up at smaller threshold $\beta$, see Fig.~\ref{fig:betaR}.

\subsection{Curvature-induced changes of nucleation}
\label{sec:resultsCurvature}

Next, we analyze the nucleation of excitation waves on the torus inside and outside, respectively.

\paragraph{Torus inside}

\begin{figure}[b!]
\centering
\includegraphics[width=0.6\textwidth]{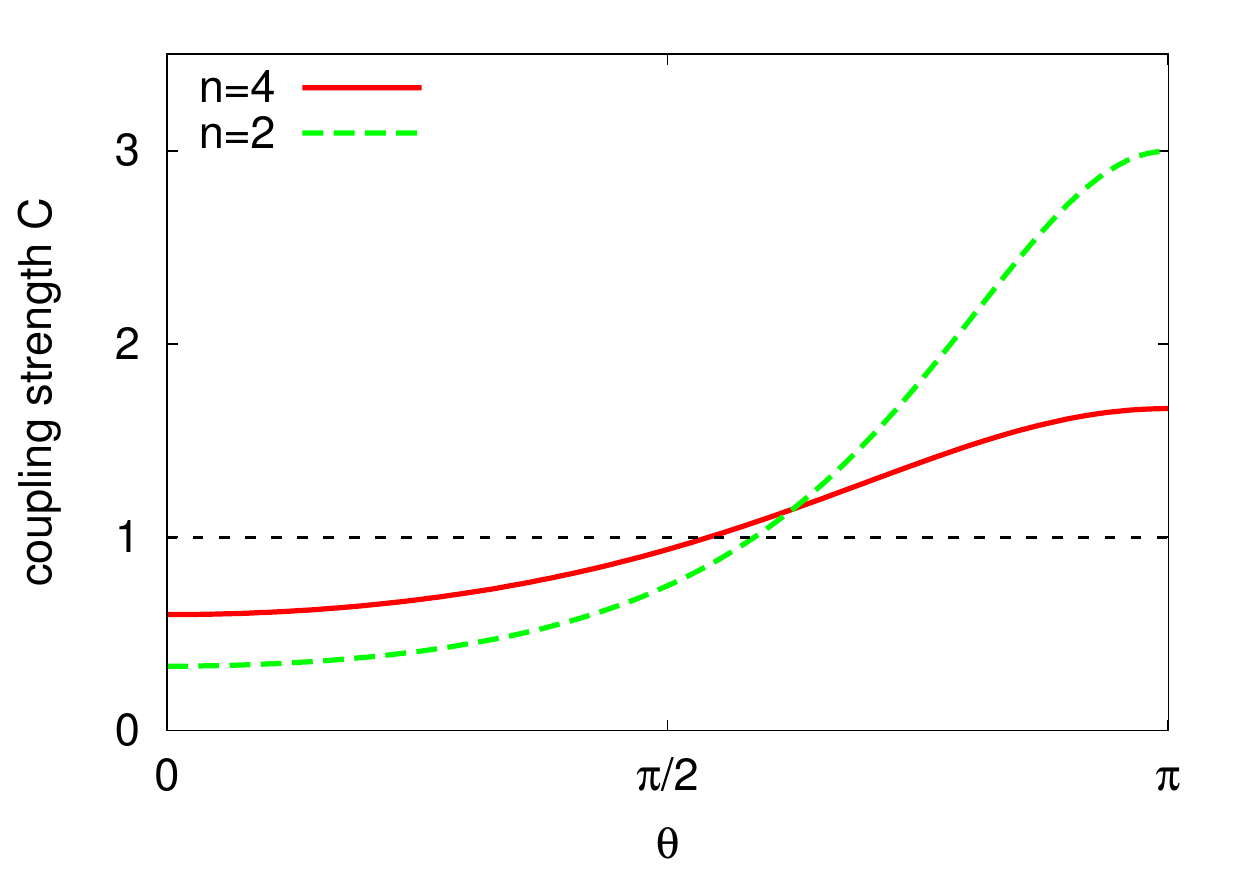}
\caption{Coupling strength $C$ as a function of common toroidal angular variable $\theta$, where the index $c$ denotes common torus coordinates, for two tori with different Gaussian curvature $\Gamma$ with $n=\frac{R}{r}$.}
\label{fig:coupling}
\end{figure}

The inside branches of the critical nucleus (blue dashed) in Fig.~\ref{fig:bifDiaResults} are to the right of the reference curve (rotor boundary $ \partial R$ on flat surfaces).
The larger the size $S$ of the critical nucleus is, the stronger is the shift towards larger threshold $\beta$. On the more strongly curved torus (torus 2), the branch of the critical nucleus is shifted further.\\
Thus, on the torus inside, there exist critical nuclei in the subexcitable parameter regime, where on flat surfaces all wave segments retract and vanish, cf.~Fig.~\ref{fig:bifDiaControl}.\\
A qualitiative explanation of this behaviour can be given by the relation of the Gaussian curvature $\Gamma$ at the center of mass of the critical nucleus ($\theta=\pi$) and at the open ends of the critical nucleus. Mathematically, the Gaussian curvature is described by the Laplace-Beltrami operator Eq.~(\ref{eq:LB}) in torus coordinates \cite{KUE05}. A torus admits a global isothermal orthogonal coordinate system, so-called toroidal coordinates $(\theta_i,\tilde{\varphi})$, that is, coordinates, where the metric is locally conformal to the Euclidean metric. The Laplace-Beltrami operator Eq.~(\ref{eq:LB}) given in toroidal coordinates is
\begin{eqnarray}
 \Delta_{LB}&=&\frac{(\cosh\eta-\cos\theta_i)^2}{a^2}(\frac{\partial^2u}{\partial\theta_i^2}+\frac{\partial^2u}{\partial\tilde{\varphi}^2}),
\end{eqnarray}
where $a=({R^2-r^2})^{\frac{1}{2}}$ is a  measure for the scaling of the space, $\eta=\arcoth [R/({R^2-r^2})^{\frac{1}{2}}]$ is a measure for the relation between the major radius $R$ and the minor radius $r$ and $\tilde{\varphi}=\varphi\sinh\eta$. The derivation can be found in the Appendix.\\
Introducing an effective  coupling strength $C(\theta_i)=(\cosh\eta-\cos\theta)^2/a^2$, a torus can mathematically be interpreted as a flat medium with a spatial coupling being a function only of the location $\theta$ ($\theta_i$ can be expressed in terms of $\theta$, see Appendix), i.e. $\tilde{D}(\theta)=DC(\theta)$. 
This is similar to effective geometric potentials arising from curved surfaces in hard condensed-matter physics \cite{KOS13}.
The coupling strength $C(\theta)$ is strictly monotonically increasing from the torus outside ($\theta=0$) to the torus inside ($\theta=\pi$), see Fig.~\ref{fig:coupling}. For more strongly curved tori, the gradient of $C(\theta)$ is larger.\\
The effective coupling strength $C(\theta)$ of the inside critical nucleus is larger at the center of mass than at the open ends. Thus, the resultant diffusion perpendicular to the propagation direction is directed towards the open ends. This counteracts the retraction of the open ends in the parameter regime which is subexcitable in a flat medium. Larger critical nuclei reach over a region of larger difference in effective coupling strength, thus the shift towards larger threshold $\beta$ is stronger.\\
The propagation velocity $c$ of the center of mass of the inside critical nucleus is similar to the propagation velocity of the ring-shaped traveling waves at the torus inside, see Fig.~\ref{fig:velocity}.

\paragraph{Torus outside}

On the torus outside, we find that, under certain conditions, unstable critical nuclei bifurcate into stable propagating wave segments (green solid line in Fig.~\ref{fig:bifDiaResults}). Furthermore, we find stable oscillating wave segments,
whose size oscillates periodically in a self-sustained way. This striking bifurcation pattern will be explained in Sect.~\ref{sec:resultsStabilisation}.\\
The outside branches (green) in Fig.~\ref{fig:bifDiaResults} are to the left of the flat reference branch $\partial R$ (black dashed). 
 On more strongly curved tori, the branch of the critical nucleus is further shifted.\\
The coupling strength relation between the torus inside and outside also explains this behaviour. As the coupling strength $C(\theta)$ at the center of mass of the outside critical nucleus is smaller than at the open ends, the resultant diffusion perpendicular to the propagation direction is directed towards the center of mass, which enhances the retraction of the open ends.\\ 
On the torus outside, critical nuclei with increasing size $S$ are found at decreasing threshold $\beta$. 
This is distinct from the inside nucleation branch and the flat reference branch: 
the larger the size $S$ of the critical nucleus is, the larger is the difference between the coupling strength at the center of mass and the coupling strength at the open ends. Thus, larger critical nuclei at the torus outside are shifted to smaller threshold $\beta$, whereas on the torus inside larger critical nuclei are shifted to larger threshold $\beta$.

Critical nuclei with small size $S$ extend over an area with almost constant coupling strength $C$, see Fig.~\ref{fig:coupling}. This supports the assumption that the branches of the inside and outside critical nucleus with small wavesize $S$ in Fig.~\ref{fig:bifDiaResults} lie close to the flat reference branch. This implies that the outside nucleation branch (green solid in Fig.~\ref{fig:bifDiaResults}) at small wavesize $S$ terminates in a saddle-node bifurcation, and the unstable outside nucleation branch coalesces with the flat reference branch; this could, however, not be resolved numerically.

\begin{figure}
\centering
\subfigure{\includegraphics[width=0.8\textwidth]{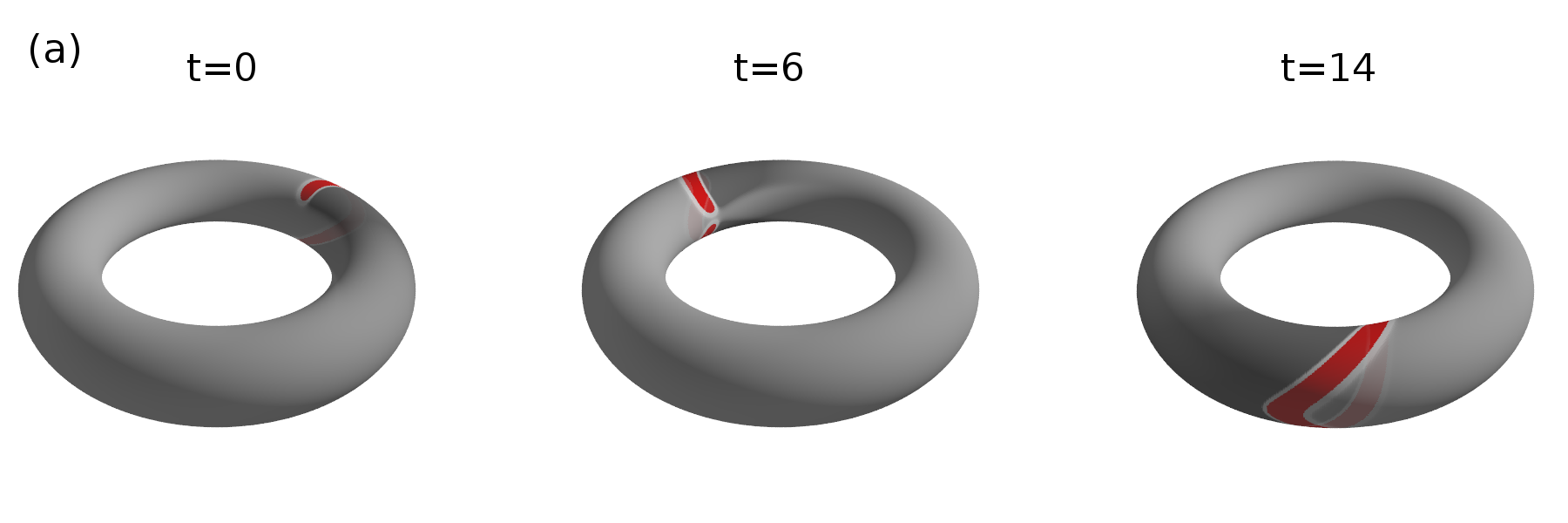}}
\subfigure{\includegraphics[width=0.8\textwidth]{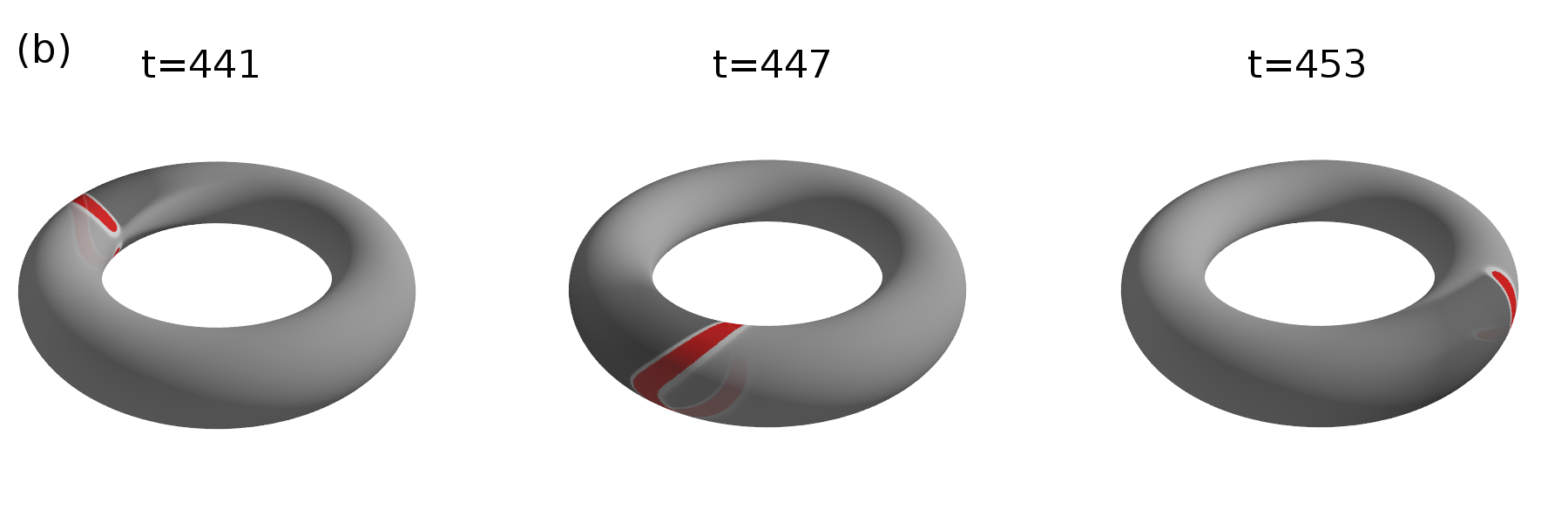}}
\subfigure{\includegraphics[width=0.8\textwidth]{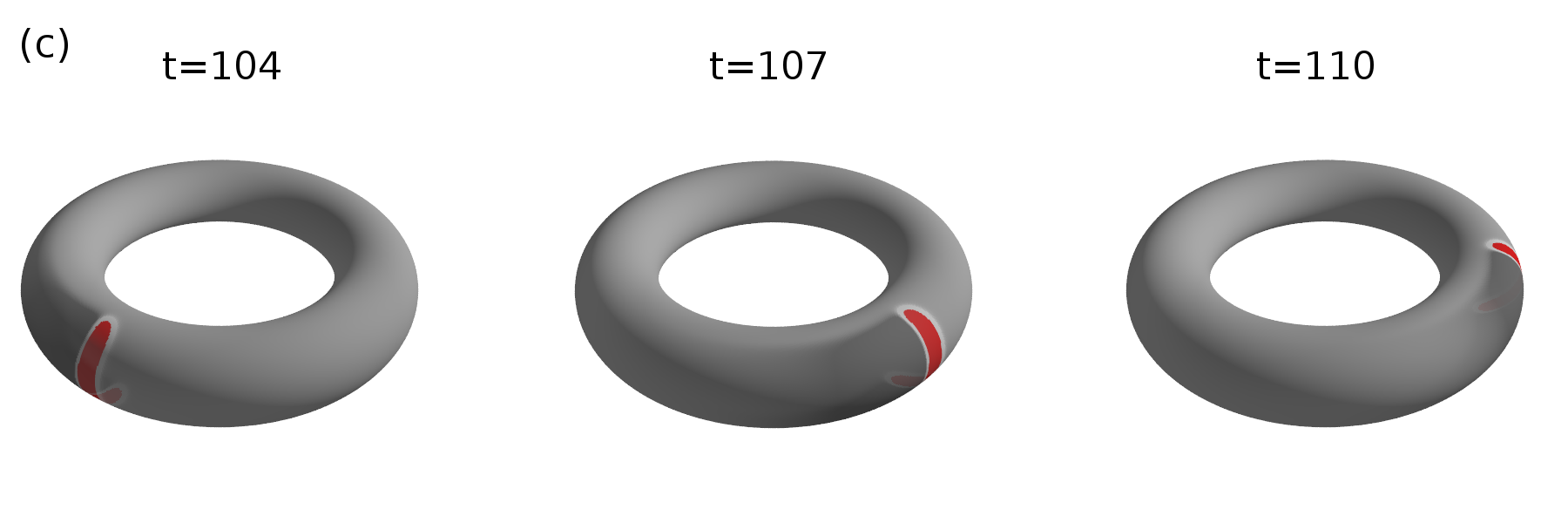}}
\subfigure{\includegraphics[width=0.8\textwidth]{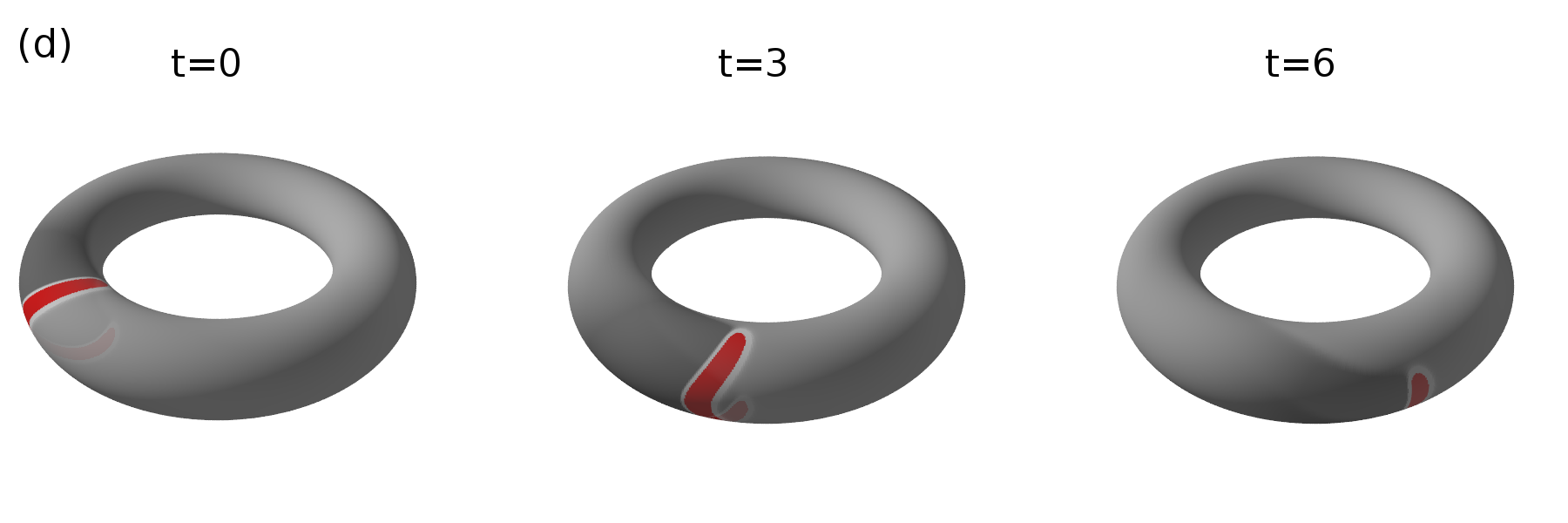}}
\caption{Snapshots of wave segments propagating counter-clockwise on a torus with minor curvature radius $r=\frac{20}{2\pi}$ and major curvature radius $R=\frac{80}{2\pi}$  computed from Eqs.(~\ref{eq:fhn1}),(~\ref{eq:fhn2}) with  $D=0.12$ and $\varepsilon=0.36$. (a) Wave segment growing towards ring-shaped traveling wave, $\beta=1.315$. (b) Oscillating wave segment, $\beta=1.321476$. (c) Stable propagating wave segment, $\beta=1.325$. (d) Wave segment shrinking towards homogeneous steady state, $\beta=1.33$. 
Movies are available from (a)\,https://www.youtube.com/watch?v=4mz7vgOvpeg (b)\,https://www.youtube.com/watch?v=\_4CrldTDsRc (c)\,https://www.youtube.com/watch?v=zZU7SaxL8Ps (d)\,https://www.youtube.com/watch?v=fw4rPG\_Lg\_4}
\label{fig:shrink}
\end{figure}

\subsection{Curvature-induced stabilisation}
\label{sec:resultsStabilisation}

\begin{figure}
\centering
 \includegraphics[width=.49\textwidth]{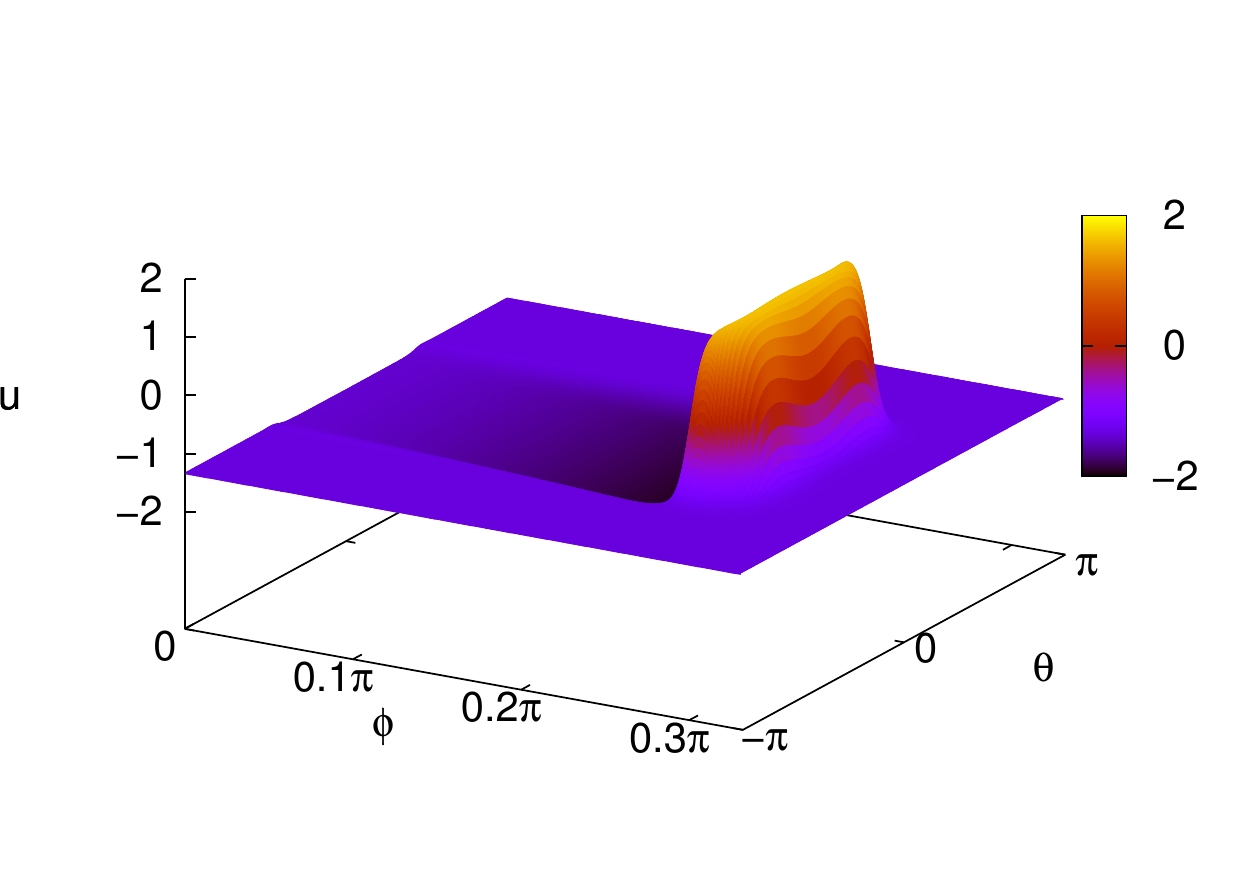}
 \caption{Snapshot of a stable propagating wave segment on the torus outside: activator concentration $u(\varphi,\theta$) on a torus with minor curvature radius $r=\frac{20}{2\pi}$ and major curvature radius $R=\frac{80}{2\pi}$ computed from Eqs.(~\ref{eq:fhn1}),(~\ref{eq:fhn2}) with $\beta=1.324$, $D=0.12$ and $\varepsilon=0.36$.}
 \label{fig:profStable}
\end{figure}

\begin{figure}
 \centering
 \subfigure[Minimal size]{\includegraphics[width=.49\textwidth]{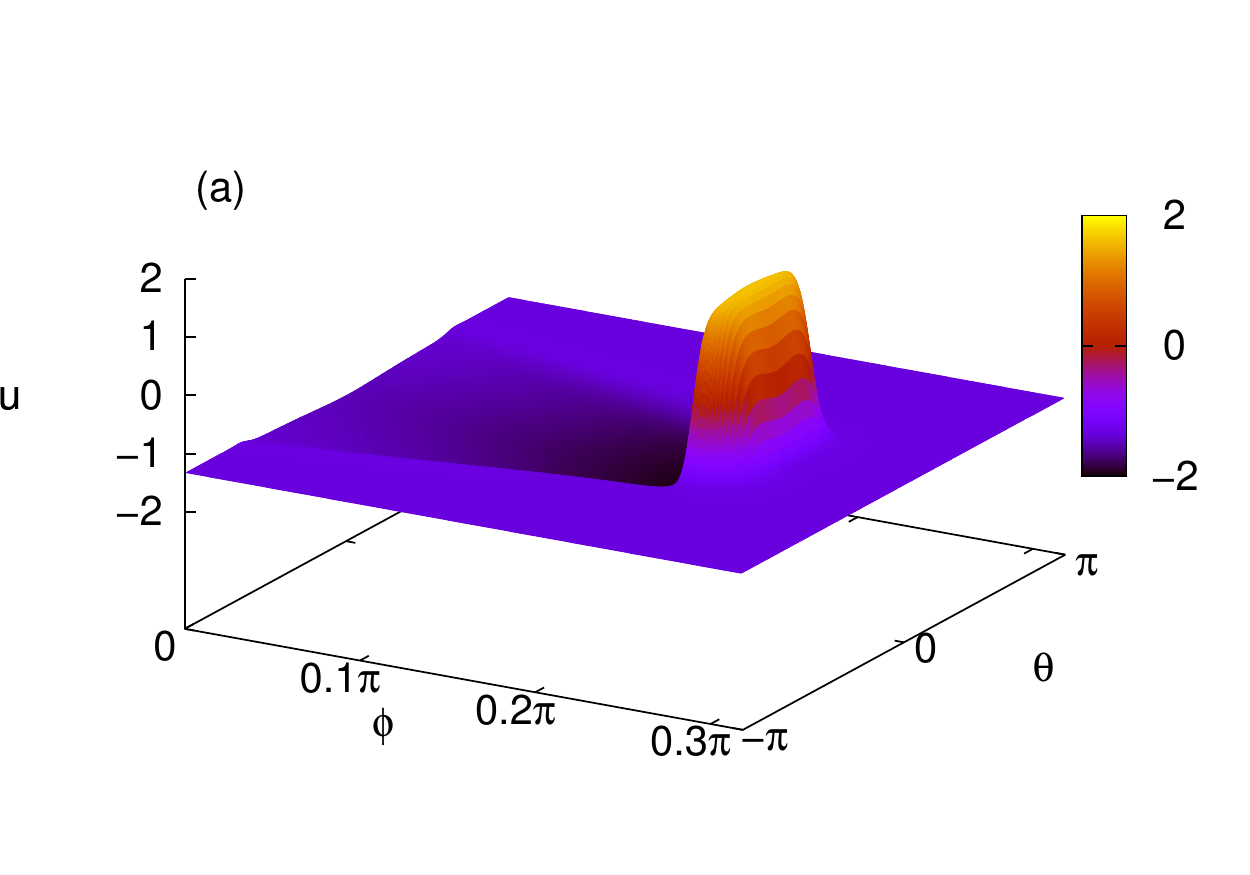}}
 \subfigure[Maximal size]{\includegraphics[width=.49\textwidth]{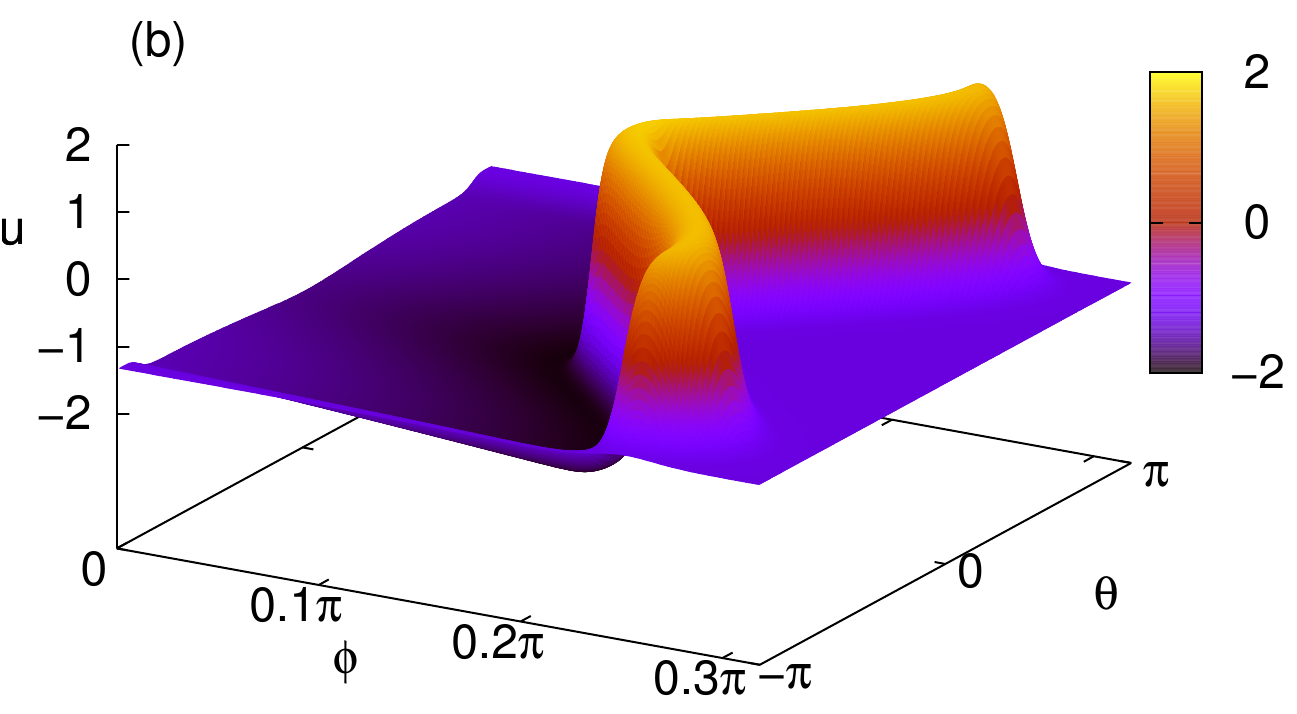}}
 \caption{Snapshots of oscillating wave segment on the torus outside: same plots as in Fig.~\ref{fig:profStable} 
with $\beta=1.321476$, $D=0.12$ and $\varepsilon=0.36$. (a) $t=478.5$ (minimum size), (b) $t=492.5$ (maximum size).}
 \label{fig:profOscillating}
\end{figure}

Depending upon the excitation parameter $\beta$, different space-time patterns occur
(Fig.~\ref{fig:shrink}). Localized wave segments may either grow to become stable ring waves (Fig.~\ref{fig:shrink}(a)), or they may shrink and vanish (Fig.~\ref{fig:shrink}(d)). Additionally, as an effect of the curved surface, 
on the torus outside, we find stable propagating localized wave segments, see Fig.~\ref{fig:shrink}(c). 
Furthermore, we find stable oscillating wave segments, whose size oscillates periodically, see Fig.~\ref{fig:shrink}(b).
In Fig.~\ref{fig:profStable}, we show the activator profile of a stable wave segment propagating with a stationary profile, and in Figs.~\ref{fig:profOscillating}(a) and (b) we show snapshots of the activator profile of an oscillating wave segment at its minimum size $S$ and at its maximum size $S$, respectively.\\
The existence of stable wave segments on surfaces with positive Gaussian curvature can be qualitatively explained with the help of the space-dependent effective coupling strength $C$, as discussed in Sect.~\ref{sec:resultsCurvature}. The open ends of a stable wave segment on the torus outside lie in an area of the torus where the coupling strength $C$ is larger than the coupling strength at $\theta=0$, where the center of mass of the wave segment is located. Thus the resultant effective diffusion perpendicular to the propagation direction caused by curvature is directed towards the center of mass of the wave segment. The larger the size $S$ of the perturbation is, the stronger is this effect.
At the same time, in the excitable parameter regime (see Fig.~\ref{fig:bifDiaControl}), small perturbations grow in length. If these two effects are balanced, we find stable propagating wave segments.
In Fig.~\ref{fig:bifDiaResults}, we show the branch of stable wave segments (green solid). Perturbations with size $S$ larger than the stable wave segments (and smaller than ring waves) shrink, as the difference in effective coupling strength between the center of mass and the open ends is large.
Perturbations with size $S$ smaller than the stable wave segments and larger than the small outside critical nucleus (which is not shown in Fig.~\ref{fig:bifDiaResults} but supposed to lie close to the flat reference branch, see Sect.~\ref{sec:resultsCurvature}) grow, as the difference in coupling strength between the center of mass and the open ends is small.

In Fig.~\ref{fig:velocity}, we show the branch of the propagation velocity $c$ at the center of mass of the stable wave segments and the stable oscillating wave segments (green solid). Furthermore, we show a hypothetical branch, the related propagation velocity at the torus inside (green dashed).

It is impossible that the stable wave segments grow to ring-shaped traveling waves (without enlarging their geodesic curvature), as the hypothetical propagation velocity at the torus inside is smaller than the minimal velocity $c_{min}$ (black solid line in Fig.~\ref{fig:velocity}).

For decreasing threshold $\beta$, the hypothetical propagation velocity at the torus inside of the stable outside wave solution (green dashed) accelerates, whereas the minimum velocity $c_{min}$ (black solid) slows down.
At the intersection point of these two branches, the stable wave solution bifurcates into a stable oscillating wave segment and an unstable critical nucleus.\\
The stable oscillating wave segments grow in length, until the propagation velocity $c$ of the open ends falls below the minimum velocity $c_{min}$. The open ends become unstable and decrease in width. Although they continue growing in length, after the minimum velocity $c_{min}$ is reached, the open ends asymptotically vanish.

\section{Conclusions}
\label{sec:conclusion}
Spreading depression (SD) is a pathological dysfunction of brain activity that occurs, e.g., during migraine or stroke. It appears as spatially localized wave segments propagating in the cortex \cite{DAH09a}. Despite substantial progress in the understanding of SD, the inherent processes are still incompletely known. Here, we study the influence of the curvature of the cortex on SD.

For this purpose, nucleation and propagation of spatially localized reaction-diffusion waves are investigated on the surface of a torus. These unstable structures, which represent critical nuclei, are stabilized by an internal feedback control Eq.~\ref{tp-b7-eq:gconstraint}. We confine attention to the case that the center of mass of the critical nuclei is pinned on the torus inside or outside, respectively.

We have shown that negative Gaussian curvature (torus inside) causes a shift of the nucleation branch to larger threshold $\beta$, i.e., there exist critical nuclei in a parameter regime that is subexcitable on flat surfaces.  In view of SD waves in the cortex, this might indicate that SD is more likely to be initiated in areas with negative Gaussian curvature.

In addition, we have made the surprising discovery that curvature can induce a change of stability, i.e., on the torus outside we have found stable propagating localized wave segments, as well as wave segments periodically oscillating in size.

These findings are qualitatively explained by an effective coupling strength, which can be found as an equivalent mathematical description of the Gaussian curvature on surfaces that admit isothermal coordinates.

Furthermore, we have reviewed the behaviour of ring-shaped wave solutions (autowaves), which were first described by V.~A.~Davydov in 2003 \cite{DAV03}. In the context of critical propagation effects, we have confirmed that the propagation boundary of ring-shaped waves, constituting break-up on the torus inside, is caused by a saddle-node bifurcation, where the fast wave branch collides with the slow wave branch. Thereby, the propagation velocity of ring-shaped waves is compared to a semianalytically calculated minimum velocity $c_{min}$ Eq.~\ref{eq:cmin}.

\section*{Appendix}
\label{sec:Appendix}

\subsection*{Minimum propagation velocity}

It is well known that wave propagation becomes impossible if the propagation velocity of a traveling wave falls below a critical value $c_{min}$ \cite{ZYK84}. Here, we consider the case of a traveling wave on a curved surface, which leads to slowing down of the wave by negative geodesic curvature of the wave front.
The geodesic curvature of a wavefront is the curvature of the front projected in its tangential plane. This should not be confused with the Gaussian curvature $\Gamma$ of a surface, which is the product of the two principal curvatures, i.e.,
\begin{eqnarray}
\Gamma&=&\frac{\cos\theta}{r(R+r\cos\theta)}
\end{eqnarray}
in common torus coordinates $(\theta,\varphi)$.
\\
The FHN model~(\ref{eq:fhn1}),(\ref{eq:fhn2}) in one spatial dimension for $1<\beta<\sqrt{3}$ and sufficiently small $\varepsilon$, has a stable fast wave solution and an unstable slow wave solution corresponding to homoclinic orbits of the related ODE problem Eqs.~(\ref{eq:fhn1comoving}),(\ref{eq:fhn2comoving}), see \cite{KRU97}. There exists a critical line in the ($\varepsilon,\beta$) space, at which the fast wave branch is connected to the slow wave branch. For values of $\beta$ or $\varepsilon$, respectively, above this critical line, propagation of traveling waves is not possible. The values of $\varepsilon$ on this critical line are denoted by $\varepsilon_{cr}$, the corresponding critical propagation velocity is $c_{cr}$.

Here we show the derivation of the analytical dependency between the minimum velocity $c_{min}$ and the critical time scale separation $\varepsilon_{cr}$ and the critical propagation velocity $c_{cr}$.

As described in \cite{ZYK84}, the propagation of slightly curved fronts (curvature radius $R\ll$ rising front width $L$) can be approximated by

\begin{eqnarray}
 \label{eq:fhn1comovingadv} c\frac{\partial u}{\partial
  \xi} &=& 3u - u^3 -v + D \frac{\partial^2 u}{\partial^2 \xi}  +\frac{D}{R}\frac{\partial u}{\partial\xi}, \\\label{eq:fhn2comovingadv}
  c\frac{\partial v}{\partial \xi} &=& \varepsilon( u+ \beta),
\end{eqnarray}
where the FHN model has been transformed to the co-moving coordinate $\xi=x+ct$, with propagation velocity $c$.

This can be written as
\begin{eqnarray}
 \label{eq:fhn1comovingadv2} (c-\frac{D}{R})\frac{\partial u}{\partial
  \xi} &=& 3u - u^3 -v + D \frac{\partial^2 u}{\partial^2 \xi},  \\\label{eq:fhn2comovingadv2}
 c\frac{\partial v}{\partial \xi} &=& \varepsilon( u+ \beta).
\end{eqnarray}

Introducing rescaled parameters $c^*$ and $\varepsilon^*$ 
\begin{eqnarray}
 c^*&=&c-\frac{D}{R},\\
 \varepsilon^*&=&\varepsilon\frac{c^*}{c},
 \label{eq:epsstar}
\end{eqnarray}
yields
\begin{eqnarray}
 c^*\frac{\partial u}{\partial\xi} &=& 3u-u^3-v+D\frac{\partial^2 u}{\partial^2
  \xi},\label{eq:fhn1star}\\
c^*\frac{\partial v}{\partial\xi} &=& \varepsilon^*(u+\beta), 
\label{eq:fhn2star}
\end{eqnarray}
which has the same form as the FHN model in one spatial dimension.

Thus, for $c^*=c_{cr}$ and $\varepsilon^*=\varepsilon_{cr}$, the homoclinic solution of Eqs.~(\ref{eq:fhn1star}),(\ref{eq:fhn2star}) corresponds to the connection between the fast wave branch and the slow wave branch. Hence, the minimal velocity $c_{min}$ of curved fronts can be derived from Eq.~(\ref{eq:epsstar}) by setting $c^*=c_{cr}$ and $\varepsilon^*=\varepsilon_{cr}$. This yields
\begin{eqnarray}
 c_{min}&=& \frac{\varepsilon}{\varepsilon_{cr}}c_{cr}.
\end{eqnarray}

\subsection*{Toroidal coordinates} 

A parametrization  $f:\{\alpha^i\}\mapsto\{x^j\}$ gives the Laplace-Beltrami operator in curvilinear coordinates 

\begin{eqnarray}
\boldsymbol{\Delta}_{LB}&=&\sum_{i,k}\frac{1}{\sqrt{g}}\frac{\partial}{\partial\alpha^i}\left(g^{ik}\sqrt{g}\frac{\partial}{\partial\alpha^k}\right),
\label{eq:LBappendix}
\end{eqnarray}
where $\textbf{G}$ is the metric tensor with matrix elements $g_{ik}$, which is the product of the transposed Jacobian matrix of $f$ multiplied with the Jacobian matrix of $f$, and $g=\Det\textbf{G}$, see \cite{KUE05}. The single components of the metrical tensor thus are the scalar product

\begin{eqnarray}
g_{ik}&=&\sum_j\frac{\partial f_j}{\partial\alpha^i}\frac{\partial f_j}{\partial\alpha^k}=:
\left\langle\frac{\partial f}{\partial\alpha^i}\mid\frac{\partial f}{\partial\alpha^k}\right\rangle.
\end{eqnarray}
A parametrization $f$ is isothermal, if the derived coordinate system is orthogonal and conformal.
In two spatial dimensions, a parametrization

\begin{eqnarray}
f:
(\alpha^1,\alpha^2)
&\mapsto&
\left(
\begin{array}{c}x\\y\\z \end{array}
\right)
\end{eqnarray}
is orthogonal, if the scalar product of the basis vectors for $i\neq k$ equals zero,
\begin{eqnarray}
\left\langle\frac{\partial f}{\partial\alpha^i}|\frac{\partial f}{\partial\alpha^k}\right\rangle&=&0.
\end{eqnarray}
The condition for conformal mapping is
\begin{eqnarray}
 \left\langle\frac{\partial f}{\partial\alpha^i}|\frac{\partial f}{\partial\alpha^i}\right\rangle&=& \left\langle\frac{\partial f}{\partial\alpha^k}|\frac{\partial f}{\partial\alpha^k}\right\rangle,
\end{eqnarray}
see \cite{KUE05}.\\
This yields the following form of the Laplace-Beltrami operator

\begin{eqnarray}
\boldsymbol{\Delta}_{LB}&=&\sum_{i,k}\frac{1}{\sqrt{g}}\frac{\partial}{\partial\alpha^i}\frac{\partial}{\partial\alpha^k}\delta^{ik}=\frac{1}{\sqrt{g}}\nabla^2.
\label{eq:LBiso}
\end{eqnarray}
To derive a global isothermal coordinate system for the surface of a torus, we start from the parametrization \cite{MOO71}
\begin{eqnarray}
(\theta_i,\varphi)
&\mapsto&
\left(
\begin{array}{c}
\frac{a\sinh\eta\cos\varphi}{\cosh\eta-\cos\theta_i}\\\frac{a\sinh\eta\sin\varphi}{\cosh\eta-\cos\theta_i}\\\frac{a\sin\theta_i}{\cosh\eta-\cos\theta_i}
\end{array}
\right)
=
\left(
\begin{array}{c}x\\y\\z\end{array}
\right),
\label{eq:parametrization}
\end{eqnarray}
where $a>0$ is a scaling factor of space and $\eta>0$ is a measure for the ratio of major curvature radius $R$ and minor curvature radius $r$. As can easily be proved, these are orthogonal coordinates.\\
As
\begin{eqnarray}
g_{\theta_i\theta_i}&=&\frac{a^2}{(\cosh\eta-\cos\theta_i)^2}\nonumber
\end{eqnarray}
and 
\begin{eqnarray}
g_{\varphi\varphi}&=&\frac{a^2\sinh^2\eta}{(\cosh\eta-\cos\theta_i)^2},\nonumber
\end{eqnarray}
$g_{\theta_i\theta_i}\neq g_{\varphi\varphi}$, thus the parametrization ~(\ref{eq:parametrization}) is not conformal. Introducing the variable 
\begin{eqnarray}
\tilde\varphi&=&\varphi\sinh\eta\nonumber
\end{eqnarray}
yields
\begin{eqnarray}
g_{\tilde{\varphi}\tilde{\varphi}}=g_{\theta_i\theta_i}=\sqrt{g}&=&\frac{a^2}{(\cosh\eta-\cos\theta_i)^2}.
\label{eq:g}
\end{eqnarray}
Thus the Laplace-Beltrami operator in isothermal torus coordinates reads
\begin{eqnarray}
 \boldsymbol{\Delta}_{LB}&=&\frac{(\cosh\eta-\cos\theta_i)^2}{a^2}\left(\frac{\partial^2u}{\partial\theta_i^2}+\frac{\partial^2u}{\partial\tilde\varphi^2}\right).
\label{eq:LBiso2}
\end{eqnarray}
To obtain the dependencies of $a$ and $\eta$ upon the major curvature radius $R$ and the minor curvature radius $r$, which are parameters of the common parametrization
\begin{eqnarray}
(\theta,\varphi)
&\mapsto
\left(
\begin{array}{c}
(R+r\cos\theta)\cos\varphi\\(R+r\cos\theta)\sin\varphi\\r\sin\theta
\end{array}
\right)=
\left(
\begin{array}{c}
x\\y\\z
\end{array}
\right),
\label{eq:toroidalgewoenlich}
\end{eqnarray}
one needs to compare Eq.~(\ref{eq:toroidalgewoenlich}) with the isothermal parametrization
\begin{eqnarray}
(\theta_i,\tilde\varphi)
&\mapsto
\left(
\begin{array}{c}
\frac{a\sinh\eta\cos(\frac{\tilde\varphi}{\sinh\eta})}{\cosh\eta-\cos\theta_i}\\\frac{a\sinh\eta\sin(\frac{\tilde\varphi}{\sinh\eta})}{\cosh\eta-\cos\theta_i}\\\frac{a\sin\theta_i}{\cosh\eta-\cos\theta_i}
\end{array}
\right)
=
\left(
\begin{array}{c}
x\\y\\z
\end{array}
\right).
\label{eq:toroidaliso}
\end{eqnarray}
A necessary and sufficient condition that a point from the domain of definition of the parametrization Eq.~(\ref{eq:toroidalgewoenlich}) lies on the twodimensional surface of a torus in the Euclidian $\mathbb{R}^3$ is
\begin{eqnarray}
  \left(\sqrt{x^2 + y^2}-R\right)^2 + z^2 - r^2 &=& 0.
\label{eq:3}
\end{eqnarray}
In toroidal coordinates Eq.~(\ref{eq:toroidaliso}) yields
\begin{eqnarray}
  x^2 + y^2 + z^2 - 2a\frac{\cosh\eta}{\sinh\eta} \sqrt{x^2+y^2} + a^2 &=& 0.
\label{eq:4}
\end{eqnarray}
By comparing the coefficients one obtains from Eq.~(\ref{eq:3}) and Eq.~(\ref{eq:4})
\begin{eqnarray}
 R &=& a \frac{\cosh\eta}{\sinh\eta} = a \coth\eta,\label{eq:R}\\
 r &=& a \kw{\sinh\eta},
 \label{eq:r}
\end{eqnarray}
and the inverse relations
\begin{eqnarray}
  a &=& \sqrt{R^2-r^2},  \\
  \eta &=& \arcoth\frac{R}{\sqrt{R^2-r^2}}=\arcoth\frac{n}{\sqrt{n^2-1}},
\end{eqnarray}
where $n=\frac{R}{r}$.\\
As can be seen from the parametrizations~(\ref{eq:toroidalgewoenlich}) and ~(\ref{eq:toroidaliso}), the transformation $\tilde\varphi(\varphi)$ is
\begin{eqnarray}
 \tilde\varphi(\varphi)&=&\varphi\sinh\eta.
\end{eqnarray}
To derive the dependency between $\theta_i$ and $\theta$, the expressions $\sqrt{x^2+y^2}-R$ of both coordinate systems are compared. This yields
\begin{eqnarray}
 r\cos\theta&=&a\frac{\sinh\eta}{\cosh\eta-\cos\theta_i}-R.
\end{eqnarray}
Replacing $R$ and $r$ with Eqs.~(\ref{eq:R}) and ~(\ref{eq:r}), this yields
\begin{eqnarray}
  \theta                                &=        \arccos\left(\frac{\cosh\eta\cos\theta_i-1}{\cosh\eta-\cos\theta_i}\right)\cdot
                                            \left\{\begin{smallmatrix}
                                                    +1        &        \theta_i \ge 0        \\
                                                    -1        &        \theta_i < 0
                                                   \end{smallmatrix}
                                            \right.
\label{eq:thetavontheta}
\end{eqnarray}
The inverse function is
\begin{eqnarray}
  \theta_i                        &=        \arccos\left(\frac R r - \frac{R^2-r^2}{r(R+r\cos\theta)}\right)\cdot
                                            \left\{\begin{smallmatrix}
                                                    +1        &        \theta \ge 0        \\
                                                    -1        &        \theta < 0
                                                   \end{smallmatrix}
                                            \right.
\label{eq:theetavonteta}
\end{eqnarray}




\end{document}